\documentclass[a4paper,fleqn,usenatbib]{mnras}

\usepackage[T1]{fontenc}
\usepackage{ae,aecompl}

\usepackage{graphicx}	
\usepackage{amsmath}	
\usepackage{amssymb}	
\usepackage{url}
\usepackage{gensymb}
\usepackage{subcaption}
\usepackage{float}
\usepackage{placeins}
\usepackage{soul}

\usepackage[dvipsnames]{xcolor}

\usepackage{tikz}
\usepackage{graphicx}
\usepackage{tabularx} 

\usepackage [english]{babel}
\usepackage [autostyle, english = american]{csquotes}
\MakeOuterQuote{"}

\title[BAO peak significance]{Detection significance of Baryon Acoustic Oscillations peaks in galaxy and quasar clustering}

\author[B. Ansarinejad]{
Behzad Ansarinejad$^{1}$\thanks{E-mail: behzad.ansarinejad@durham.ac.uk} \& Tom Shanks$^{1}$
\\
$^{1}$Department of Physics, Durham University, South Road, Durham, DH1 3LE, UK\\
}

\date{Accepted XXX. Received YYY; in original form ZZZ}

\pubyear{2018}

\begin{document}
\label{firstpage}
\pagerange{\pageref{firstpage}--\pageref{lastpage}}
\maketitle

\begin{abstract}
We compare our analysis of the Baryon Acoustic Oscillations (BAO) feature in the correlation functions of SDSS BOSS DR12 LOWZ and CMASS galaxy samples with the findings of \cite{Cuesta2016}. Using subsets of the data we obtain an empirical estimate of the errors on the correlation functions which are in agreement with the simulated errors of \cite{Cuesta2016}. We find that the significance of BAO detection is the quantity most sensitive to the choice of the fitting range with the CMASS value decreasing from $8.0\sigma$ to $5.3\sigma$ as the fitting range is reduced. Although our measurements of $D_V(z)$ are in agreement with those of \cite{Cuesta2016}, we note that their CMASS $8.0\sigma$ (LOWZ $4.0\sigma$) detection significance reduces to $4.7\sigma$ ($2.8\sigma$) in fits with their diagonal covariance terms only. We extend our BAO analysis to higher redshifts by fitting to the weighted mean of 2QDESp, SDSS DR5 UNIFORM, 2QZ and 2SLAQ quasar correlation functions, obtaining a $7.6\%$ measurement compared to $3.9\%$ achieved by eBOSS DR14. Unlike for the LRG surveys, the larger error on quasar correlation functions implies a smaller role for nuisance parameters (accounting for scale-dependent clustering) in providing a good fit to the fiducial $\Lambda$CDM model. Again using only the error bars of \cite{Ata2017} and ignoring any off-diagonal covariance matrix terms, we find that the eBOSS peak significance reduces from 2.8 to $1.4\sigma$. We conclude that for both LRGs and quasars, the reported BAO peak significances from the SDSS surveys depend sensitively on the accuracy of the covariance matrix at large separations. 
\end{abstract}




\section{Introduction}
\label{sec:BOSS_intro}
The determination of the expansion history of the universe is currently
one of the primary goals of observational cosmology. The late-time
transition of the expansion rate of the universe from a deceleration to
a phase of acceleration (e.g. based on observational evidence from
supernovae; \citealt{Riess1998}; \citealt{Perlmutter1999}) in
particular, remains one of the most puzzling problems in modern physics.
Investigating this problem and exploring the nature of Dark Energy (a
hypothetical cause of the accelerated expansion rate of the universe
\citep{Peebles2003}, within the framework of $\Lambda$CDM, the current
standard cosmological model), have driven efforts to obtain robust and
high precision measurements of the cosmological expansion rate. To this
end, a great interest was sparked in exploiting large galaxy redshift
surveys in order to constrain the distance-redshift relation across a
wide range of redshifts, making use of the Baryon Acoustic Oscillation
(BAO) feature in the clustering of galaxies (e.g. \citealt{Shanks1985};
\citealt{Blake2003}; \citealt{Linder2003}; \citealt{Seo2003};
\citealt{Matsubara2004}; \citealt{Glazebrook2005}; \citealt{Donley2006};
\citealt{Sanchez2008}). 

A measurement of the BAO signature in the monopole two point correlation
function of the "Constant Stellar Mass" (CMASS) and the low-redshift
(LOWZ) galaxy samples from the Data Release 12 (DR12;
\citealt{Alam2015}) of the SDSS BOSS survey was presented by
\cite{Cuesta2016}. The CMASS and LOWZ samples are extensions to previous
SDSS LRG samples.

Here, we first present the results of our independent measurement of
the BAO feature in the DR12 CMASS and LOWZ samples. This is followed by a
comparison to results of \cite{Cuesta2016} providing an independent verification of the applied methodology, placing particular focus on the uncertainties on
the correlation functions. \cite{Cuesta2016} obtained an estimate of the uncertainties based on the covariance matrix of $1000$ BOSS DR12
simulated QPM mocks \citep{WTM2014}. In this study, we divide the data
into subsamples upon which measurements of the correlation function are performed, giving an empirical estimate of the uncertainty on the mean correlation function. Furthermore, we investigate certain aspects of the fitting procedure commonly implemented in BAO analysis studies. These include the extent of the role played by the nuisance fitting parameters in providing a good fit; effects of the choice of the fitting range on the results and a comparison between fits using the full BOSS DR12 QPM covariance matrices and their diagonal elements only. Here our main goal is to investigate the robustness of the BAO peak detection significance to variations in different aspects of the fitting procedure. Note that in this work we do not attempt to perform reconstruction and hence we simply draw comparison with the pre-reconstruction results throughout. 

At higher redshifts, BAO have also been detected in the Lyman-alpha
forest in the BOSS quasar survey at $2.1<z<3.3$ (\citealt{Slosar2013}; \citealt{Delubac2015}). As originally suggested by \cite{Sawangwit2012}, it is also possible to make accurate BAO measurements in the $z<2.2$ range using quasars as direct tracers of the matter distribution. The eBOSS survey \citep{Dawson2016} is therefore making BAO measurements via quasars using them both directly as tracers and via the Lyman-alpha forest. Here we shall use the SDSS DR5 \citep{SDSS_CF_ROSS2009}, 2SLAQ \citep{2SLAQ_cat_Croom2009}, 2QZ \citep{2QZ_cat_Croom2004} and 2QDES pilot \citep{Chehade2016} surveys to determine the level of accuracy to which the BAO scale can be measured by the previous generation of quasar surveys used as direct tracers in the $0.8<z<2.2$ redshift range. Furthermore, we combine our results with those of \cite{Ata2017} who performed BAO analysis on the eBOSS DR14 quasar sample in the same redshift range, obtaining a BAO distance measurement based on the combination of these samples.

The layout of this paper is as follows: Section~\ref{sec:Datasets}
contains a brief description of the galaxy samples along with the basic properties of the selected subsamples. In Section~\ref{sec:Methodology} we present a description of the relevant methodology involved in measuring the correlation function, error analysis and the fitting procedure. This is followed by a presentation and discussion of our results and a comparison of our findings with those of \cite{Cuesta2016} in Section~\ref{sec:Results}. In Section~\ref{sec:QSO_intro} we provide a description of the quasar samples used in our high redshift BAO analysis, followed by an outline of our applied methodology in section~\ref{sec:QSO_Methodology}. We present the results of our QSO BAO analysis in section~\ref{sec:QSO_Results}, along with the cosmological distance constraints obtained from our QSO and LRG measurements, comparing our findings with the predictions of \cite{Planck2015}. Finally, we conclude this work by providing a summary of our findings in Section~\ref{sec:Conclusions}.

\section{Datasets}
\label{sec:Datasets}

\begin{figure}
	\includegraphics[width=\columnwidth]{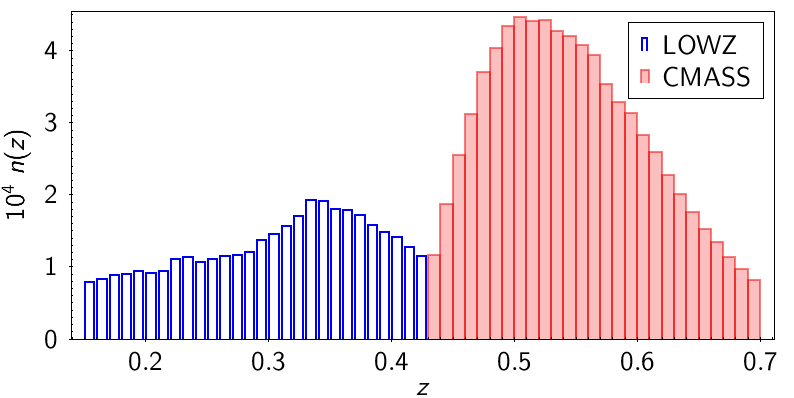}
	\caption[Redshift distribution of BOSS DR12 LOWZ and CMASS samples]{The redshift distribution of the BOSS DR12 LOWZ and CMASS samples analysed in this study, as well as in \citet{Cuesta2016}. Bins are $\Delta z=0.01$ in width.}
	\label{fig:Hist_new.pdf}
\end{figure}

In this study we first use a set of 777,202 galaxies in the redshift range $0.43<z<0.7$ from the BOSS DR12 CMASS sample, with an effective redshift of 0.57, and 361,762 galaxies in the redshift range $0.15<z<0.43$ from the DR12 LOWZ sample, with an effective redshift of 0.32. The CMASS and LOWZ samples have been limited to magnitudes of $17.5< i_{cmod}<19.9$ and $16<r_{cmod}<19.6$ respectively. Full details of the target selection criteria can be found in \cite{Reid2016} and the treatment of systematics and the relevant corrections is discussed in \cite{Ross2016}. In accordance with \cite{Cuesta2016}, the samples, mocks and random datasets were obtained from the DR12 database\footnote{\url{https://data.sdss.org/sas/dr12/boss/lss/}}. The
redshift distributions $n(z)$, of the galaxies in the DR12 CMASS and
LOWZ samples are displayed in Fig.~\ref{fig:Hist_new.pdf}. 

\begin{figure*}
	\begin{subfigure}[t]{.5\textwidth}
		\centering
		\caption{CMASS Northern Galactic Cap}
		\includegraphics[width=1\linewidth]{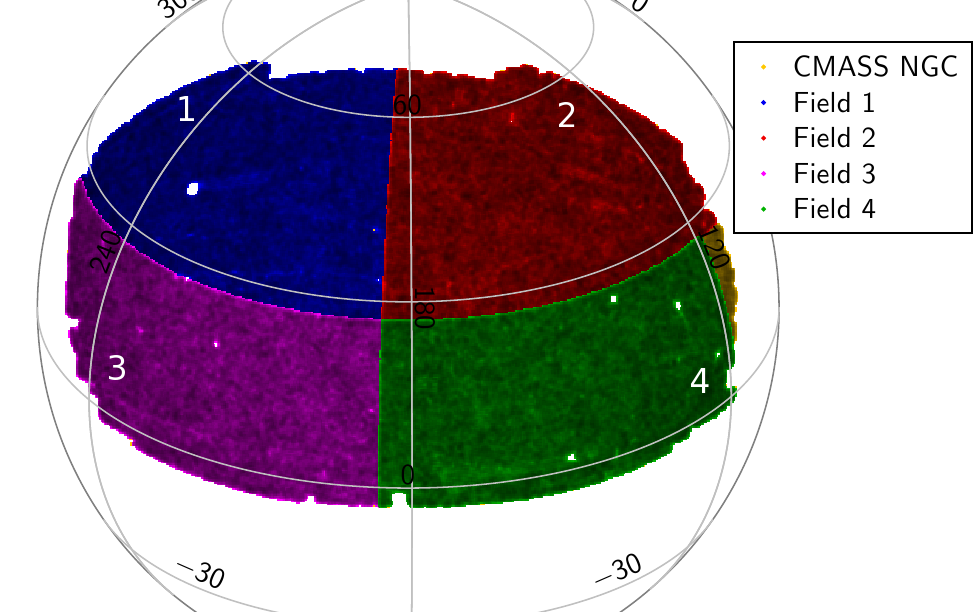}
		\label{fig:CMASS_fieldsN.pdf}
	\end{subfigure}%
	\begin{subfigure}[t]{.5\textwidth}
		\centering
		\caption{CMASS Southern Galactic Cap}
		\includegraphics[width=1\linewidth]{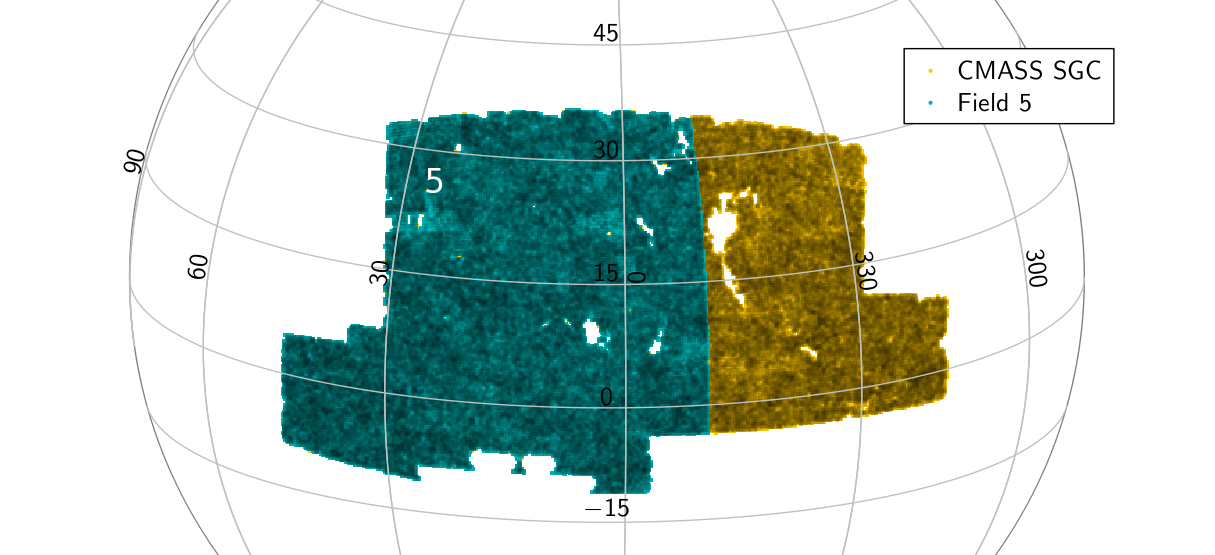}
		\label{fig:CMASS_fieldsS.pdf}
	\end{subfigure}
	\begin{subfigure}[t]{.5\textwidth}
		\centering
		\caption{LOWZ Northern Galactic Cap}
		\includegraphics[width=1\linewidth]{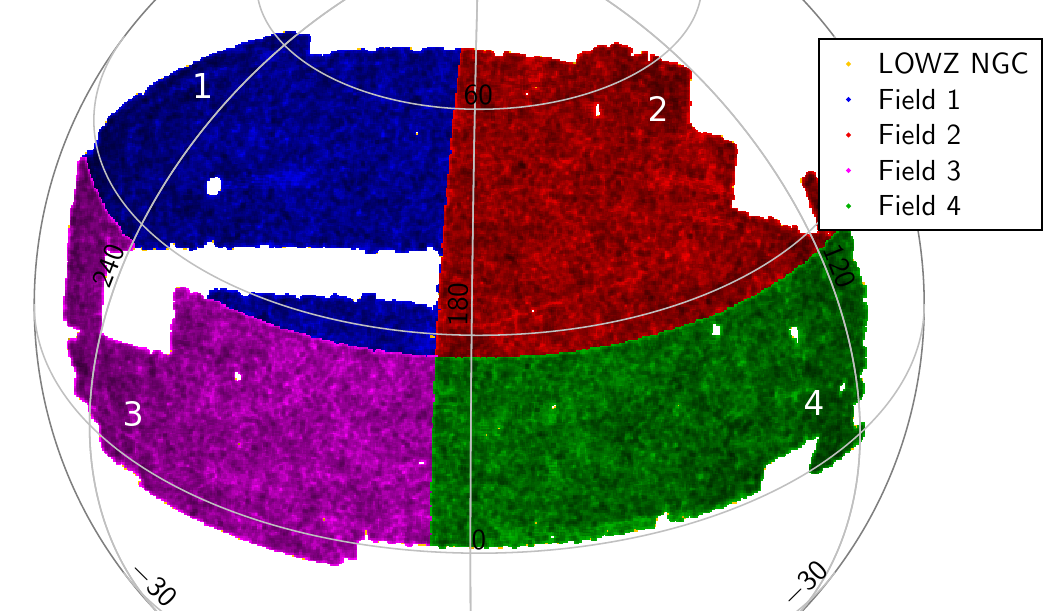}
		\label{fig:LOWZ_fieldsN.pdf}
	\end{subfigure}%
	\begin{subfigure}[t]{.5\textwidth}
		\centering
		\caption{LOWZ Southern Galactic Cap}
		\includegraphics[width=1\linewidth]{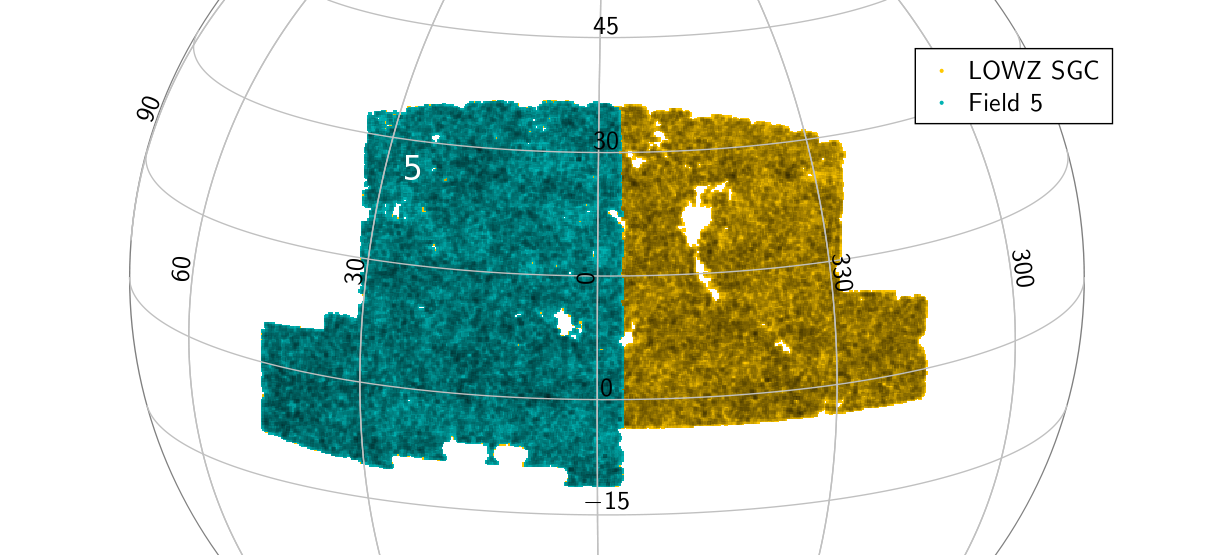}
		\label{fig:LOWZ_fieldsS.pdf}
	\end{subfigure}
	\caption[Coverage of the selected LOWZ and CMASS subsamples]{The coverage of the 5 selected fields in the Northern and Southern Galactic caps of the CMASS and LOWZ samples. The sample areas not selected are shown in yellow. The basic properties of these fields can be found in Table~\ref{tab:CMASS+LOWZ_table}.}
	\label{fig:FC_fields}
\end{figure*}

In order to obtain an empirical estimate of the uncertainties on the correlation functions, the CMASS sample is subsetted into five fields (subsamples) of equal size covering an overall area of 8487.77 $\deg^2$, about $90.5\%$ of the total effective sample area (9376.09 $\deg^2$). The LOWZ sample is similarly divided into five equally sized fields covering 7294.87 $\deg^2$, roughly $87.5\%$ of the total sample area (8337.47 $\deg^2$). Initially, dividing the samples into five fields was deemed sufficient in order to produce an estimate of the uncertainties to a reasonable degree of accuracy. However, as demonstrated in later sections, the precision of the empirical estimate of uncertainties can be further improved by using a larger number of subsamples. The positions of all selected fields are illustrated in Fig.~\ref{fig:FC_fields}, with Table~\ref{tab:CMASS+LOWZ_table} providing a description of the basic properties of the selected fields. Once the correlation function for each field is obtained, a mean correlation function is calculated and is taken to represent the correlation function of the sample, using the standard error on the mean as an estimate of the uncertainty.

\begin{table}
	\centering
	\caption[Basic properties of the LOWZ and CMASS subsamples]{The basic properties of the 5 chosen fields (shown in Fig.~\ref{fig:FC_fields}) in the CMASS and LOWZ samples.}
	\label{tab:CMASS+LOWZ_table}
	\begin{tabular}{ccccc} 
		\multicolumn{5}{c}{\textbf{CMASS}} \\ 
		\hline
		Field & Ra\degree & Dec\degree & Area ($\deg^2$) & Number of galaxies\\
		\hline
		1 & >185 & >27 & 1703 & 142,636\\
		2 & <185 & >27 & 1686 & 141,706\\
		3 & >185 & <27 & 1699 & 141,847\\
		4 & 119-185 & <27 & 1698 & 137,891\\
		5 & 350-45.5 & >-11 & 1701 & 144,820\\
		\hline
		\\
		\multicolumn{5}{c}{\textbf{LOWZ}}\\
		\hline
		Field & Ra\degree & Dec\degree & Area ($\deg^2$) & Number of galaxies\\
		\hline
		1 & >185 & >27 & 1447 & 61,319\\
		2 & <185 & >27 & 1453 & 63,109\\
		3 & >185 & <27 & 1474 & 61,605\\
		4 & <185 & <27 & 1463 & 63,431\\
		5 & 357-45.5 & >-11 & 1459 & 68,057\\
		\hline
	\end{tabular}
\end{table}

In our analysis up to Section~\ref{sec:QSO_intro}, we assume the same fiducial cosmology as \cite{Cuesta2016} with $\Omega_m=0.29$, $\Omega_bh^2=0.02247$, $\Omega_\Lambda=0.71$, $\Omega_k=0$, $\Omega_\nu=0$, $h=0.7$, $w=-1$, $n_s=0.97$ and $\sigma_8=0.8$. The fiducial distances to $z=0.32$ and $0.57$ (the effective redshifts of our samples), based on our assumed cosmology are presented in Table~\ref{tab:fid_distances}.

\begin{table*}
	\centering
	\caption[Fiducial values of $r_d$, $D_A$, $H$ and $D_V$ at $z=0.32$ and $0.57$]{A summary of the fiducial distances and values of the Hubble parameter used in this work and by \citet{Cuesta2016}, computed at the effective redshifts of the LOWZ ($z=0.32$) and CMASS ($z=0.57$) samples, based on our assumed flat $\Lambda$CDM cosmological model.}
	\label{tab:fid_distances}
	\begin{tabular}{ccccccc} 
			\hline
			$r_d$ & $D_A(z=0.32)$ & $H(z=0.32)$ & $D_V(z=0.32)$ & $D_A(z=0.57)$ & $H(z=0.57)$ & $D_V(z=0.57)$ \\
			(Mpc) & (Mpc) & (km s$^{-1}$ Mpc$^{-1}$) & (Mpc) & (Mpc) & (km s$^{-1}$ Mpc$^{-1}$) & (Mpc)\\
			\hline
			147.10 & 962.43 & 82.142 & 1235.28 & 1351.13 & 94.753 & 2009.55\\
			\hline
		\end{tabular}
	\end{table*}

\section{Methodology}
\label{sec:Methodology}

\subsection{Measuring the Correlation Function}

\label{sec:Measuring CF} 

The monopole two-point correlation function (in redshift-space), $\xi(s)$, is calculated for each individual field using the CUTE\footnote{\url{http://members.ift.uam-csic.es/dmonge/CUTE.html}} algorithm described by \cite{Alonso2012}. 

To perform the measurement of the correlation function we make use of the Landy-Szalay estimator \citep{LS1993},
\begin{equation}
\xi(s)=\frac{DD(s)-2DR(s)+RR(s)}{RR(s)},
\label{eq:LS}
\end{equation}
where $DD(s)$, $DR(s)$ and $RR(s)$ are data-data, data-random and random-random pair-counts respectively.	

In our analysis we make use of the BOSS DR12 FKP-weighted \citep{FKP} randoms, and in accordance with \cite{Reid2016}, apply a weighting of $w_{tot}w_{FKP}$ to the galaxies. A full description of the constituents of $w_{tot}$ is presented in \cite{Reid2016}; in short, this weight consists of three terms which account for effects of angular systematics, fibre collisions and redshift failures. In order to facilitate direct comparison with the findings of \cite{Cuesta2016}, we sum our pair counts into 25 bins of width $8\ h^{-1}$Mpc in our calculation of the correlation functions, covering the range of $s\leqslant200 h^{-1}$Mpc in redshift space.

\subsection{Error Analysis}

\label{sec:Error Analysis}

Following the procedure proposed by \cite{Norberg2009}, the bootstrap resampling method is used to provide an estimate of the errors on the mean correlation functions of our CMASS and LOWZ samples. In total we generate $N=100$ resamplings and obtain the mean correlation function $\bar{\xi}(s)$ of these resamplings. As demonstrated by \cite{Norberg2009}, an oversampling factor of 3 appears to be optimal in improving the bootstrap recipe. Hence we calculate the mean correlation function of each resampling, $\xi_n$, based on the correlation functions of $N_r=3 \times N_{sub}$ randomly selected subvolumes (with replacement), from the original $N_{sub}=5$ subvolumes defined in Section~\ref{sec:Datasets} for the CMASS and LOWZ samples. 

A second set of errors are determined for the mean correlation functions of our samples, simply based on obtaining the standard errors on the mean. This is done using,
\begin{equation}
\sigma_{mean}=\frac{\sigma_{N_{sub}-1}}{\sqrt{N_{sub}}}=\sqrt{\frac{\sum(\xi_i-\bar{\xi})^2}{N_{sub}^2-N_{sub}}},
\label{eq:Standard_err}
\end{equation}
where $\sigma_{N_{sub}-1}$ is the standard deviation normalized to $N_{sub}-1$ (as $\sigma_{mean}$ is obtained from the same dataset reducing the number of degrees of freedom by one); $N_{sub}$ is the number of subvolumes in each sample (i.e. 5); $\xi_i$ is the correlation function of the $i$th subvolume, and $\bar{\xi}$ is the mean correlation function of the sample. 

A comparison of the estimated errors from these two different methods and the errors found by \cite{Cuesta2016} based on the covariance matrix of the DR12 QPM mocks is presented in Section~\ref{sec:Error Analysis Results}.

\subsection{Fitting the Correlation Function}

\label{sec:Fitting the Correlation Function}

To fit the correlation functions we follow a procedure based on the methods described in \cite{Xu2012} and \cite{Andersonetal2012}. We present a brief description of these techniques in this section. 

We use a fitting model of the form
\begin{equation}
\xi^{fit}(s)=B^2\xi_m(\alpha s)+A(s),
\label{eq:xi_fit}
\end{equation}
\begin{flushleft}
where $\xi_m$ is defined in equation~\ref{eq:xi_m}, $B^2$ is a constant term allowing for any unknown large-scale bias and $A(s)$ is given by
\end{flushleft}
\begin{equation}
A(s)=\frac{a_1}{s^2}+\frac{a_2}{s}+a_3,
\label{eq:A}
\end{equation}
where $a_{1,2,3}$ are nuisance parameters. The $A(s)$ term is included in order to marginalise over broad-band effects due to redshift-space distortions and scale-dependent bias as well as any errors made in our assumption of the fiducial cosmology. The form of the $A(s)$ term was chosen by \cite{Xu2012} due to its simplicity and was further justified in that work by comparing it to various alternatives and demonstrating that it performs optimally in providing a good fit. We can obtain distance constraints by finding the optimum value of the scale dilation parameter $\alpha$. This parameter provides a measure of any isotropic shifts in the position of the BAO peak in the data compared to the fiducial model, due to non-linear structure growth. This term is defined as 
\begin{equation}
\alpha=\frac{D_V(z)}{r_d}\frac{r_{d,fid}}{D_{V,fid}(z)},
\label{eq:alpha}
\end{equation}
where $z$ is the redshift, $r_d$ is the sound horizon at the drag epoch and $fid$ denotes the fiducial values (given in Table~\ref{tab:fid_distances}).  An $\alpha >1$ ($\alpha<1$) indicates that the BAO peak in the observed data is located at a smaller (larger) scale compared to the peak in the model. The approximate volume-averaged distance to redshift $z$ is
\begin{equation}
D_V(z)\equiv\Bigg[cz(1+z)^2\frac{D_A(z)^2}{H(z)}\Bigg]^{1/3},
\label{eq:D_V}
\end{equation} 
where $D_A(z)$ is the angular diameter distance and $H(z)$ is the Hubble parameter at redshift $z$. This "distance" is proportional to the volume-averaged dilation factors \citep{Ballinger1996} in the redshift and angular directions at a redshift $z$.

The model correlation function in equation~(\ref{eq:xi_fit}), $\xi_m$, is given by 
\begin{equation}
\xi_m(s)=\int \frac{k^2dk}{2\pi^2}P_m(k)j_0(ks)e^{-k^2a^2},
\label{eq:xi_m}
\end{equation}
where the Gaussian term is added to damp the oscillatory transform kernel $j_0(ks)=sin(ks)/ks$ at high-$k$. Here we set $a=2 h^{-1}$Mpc, which is small enough as to not cause significant damping effects at our scales of interest.

The template power spectrum is given by
\begin{equation}
P_m(k)=[P_{lin}(k)-P_{noBAO}(k)]e^{-k^2\sum_{nl}^{2}/2}+P_{noBAO}(k),
\label{eq:P_m}
\end{equation}
where $P_{lin}$ is the linear power spectrum at $z=0$ (generated using CAMB\footnote{\url{http://cosmologist.info/camb/}}; \citealt{Lewis2000}) and $P_{noBAO}$ is the power spectrum with the BAO feature removed as described in \cite{Eisenstein1998}. The $\sum_{nl}^{2}/2$ term damps the BAO features in $P_{lin}$, accounting for the effects of non-linear structure evolution. Here we set $\sum_{nl}= 8 h^{-1}$Mpc.

The best fit values of the $B^2$, $a_1$, $a_2$ and $a_3$ fitting parameters in equation \ref{eq:xi_fit} are determined using the scipy.optimize.curve\_fit module in Python which makes use of the Levenberg-Marquardt algorithm. To obtain the optimum value of $\alpha$ we compute the $\chi^2$ goodness-of-fit indicator for fits obtained from shifting the model in the range $0.8<\alpha<1.2$ with intervals of $\Delta \alpha=0.0001$, taking the value of $\alpha$ which corresponds to the minimum $\chi^2$, ($\chi^2_{min}$). 

The $\chi^2$ function is given by
\begin{equation}
\chi^2(\alpha)=[\xi^{obs}-\xi^{fit}(\alpha)]^TC^{-1}[\xi^{obs}-\xi^{fit}(\alpha)],
\label{eq:Chi2}
\end{equation}
where $\xi^{obs}$ is the observed correlation function, $\xi^{fit}(\alpha)$ is the best fit model at each $\alpha$ and $C$ is the BOSS DR12 covariance matrix obtained from 1000 simulated QPM mocks. 

In this study we investigate potential effects on the measured value of $\alpha$ and its uncertainty based on fitting the data across various ranges, using the complete $\xi^{fit}$ model with and without the $A(s)$ nuisance parameters. Furthermore, by comparing the $\Delta\chi^2$ vs. $\alpha$ curves from fitting the $\xi^{fit}$ and $\xi^{noBAO}$ models (the latter is obtained by setting the term $P_m = P_{noBAO}$ in the model correlation function $\xi_m$), we obtain a measure of the significance at which the BAO signature is detected in the data. Here $\Delta\chi^2=\chi^2(\alpha)-\chi^2_{min}$. 

To obtain an estimate of the uncertainty in $\alpha$, we assume a Gaussian form for the probability distribution of $\alpha$,
\begin{equation}
p(\alpha_i)=\frac{e^{-\chi^2(\alpha_i)/2}}{\sum_{j}e^{-\chi^2(\alpha_j)/2}\Delta\alpha},
\label{eq:p_alpha}
\end{equation} 

\begin{flushleft}
	where the denominator is a normalization factor ensuring the distribution integrates to unity. In effect $p(\alpha_i)$ is the probability that the acoustic scale $\alpha=\alpha_i$, based on the $\chi^2$ distribution obtained from comparing the model $\xi^{fit}$ (equation~(\ref{eq:xi_fit}) with $\alpha_i$), to our observed correlation function $\xi^{obs}$. We then calculate the standard deviation of our probability distribution which serves as an estimate of the uncertainty in $\alpha$: 
\end{flushleft}
\begin{equation}
\sigma_{\alpha}= \sqrt{\langle\alpha^2\rangle - \langle\alpha\rangle^2} ;
\label{eq:sigma_alpha}
\end{equation} 

\begin{flushleft}
	here $\langle\alpha\rangle$ represents the mean of the $p(\alpha_i)$ distribution given by:  
\end{flushleft}
\begin{equation}
\langle\alpha\rangle=\sum\limits_{i} \alpha_i p(\alpha_i)\Delta\alpha,
\label{eq:mean_alpha}
\end{equation}
\begin{flushleft}
	and  
\end{flushleft}
\begin{equation}
\langle\alpha^2\rangle=\sum\limits_{i} \alpha_i^2 p(\alpha_i)\Delta\alpha.
\label{eq:mean2_alpha}
\end{equation}

The estimated uncertainty obtained from this method is equivalent to the value given by the $\Delta\chi^2$ curve at the $1 \sigma$ level (see Fig.~\ref{fig:LOWZ+CMASS+noBAO_delta_Chi2_02to2.eps}).

\section{Results and Discussion}
\label{sec:Results}

\begin{figure}
	\begin{subfigure}{\columnwidth}
		\centering
		\caption{}
		\includegraphics[width=1\linewidth]{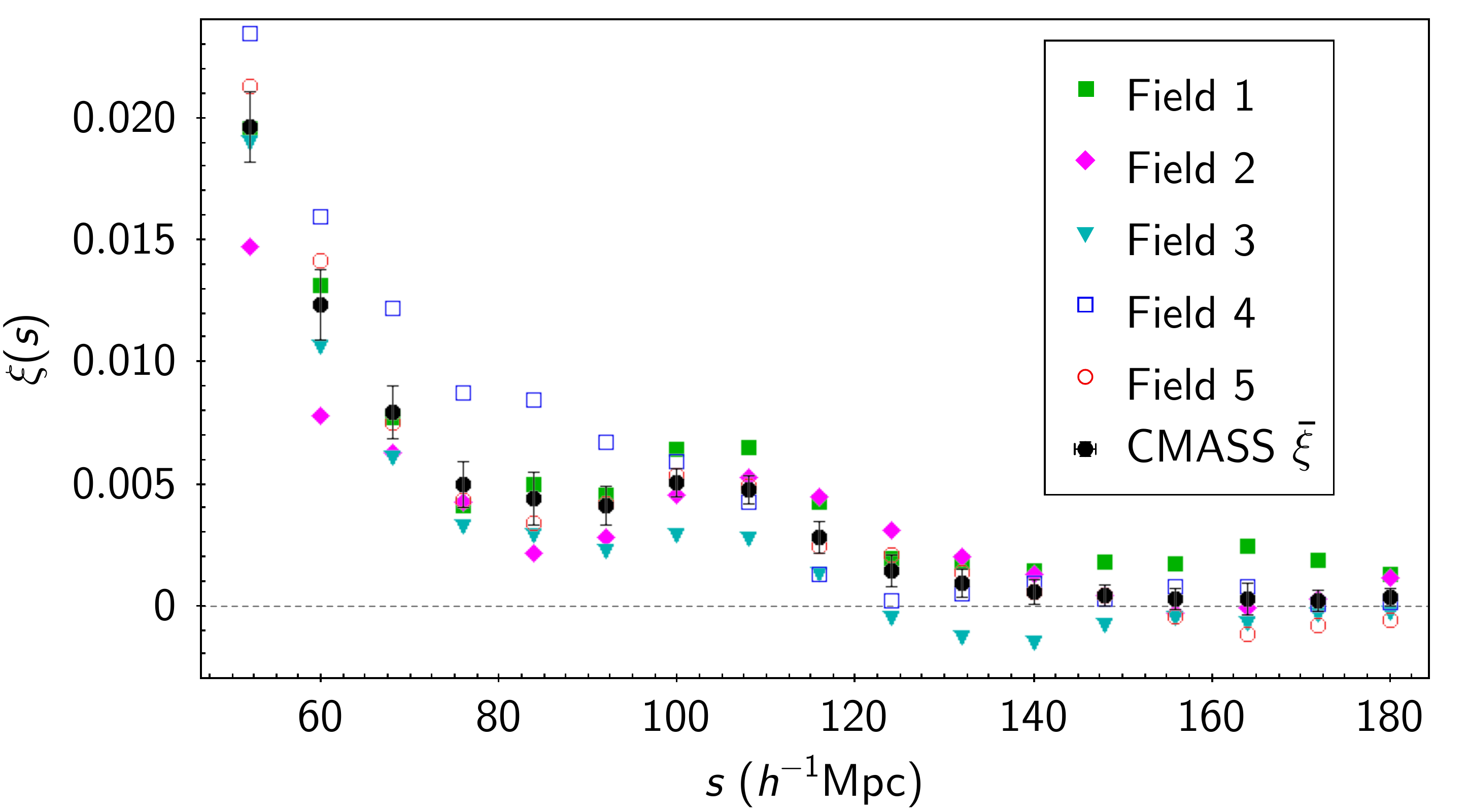}
		\label{fig:sfig_CMASS_new}
	\end{subfigure}
	\begin{subfigure}{\columnwidth}
		\centering
		\caption{}
		\includegraphics[width=1\linewidth]{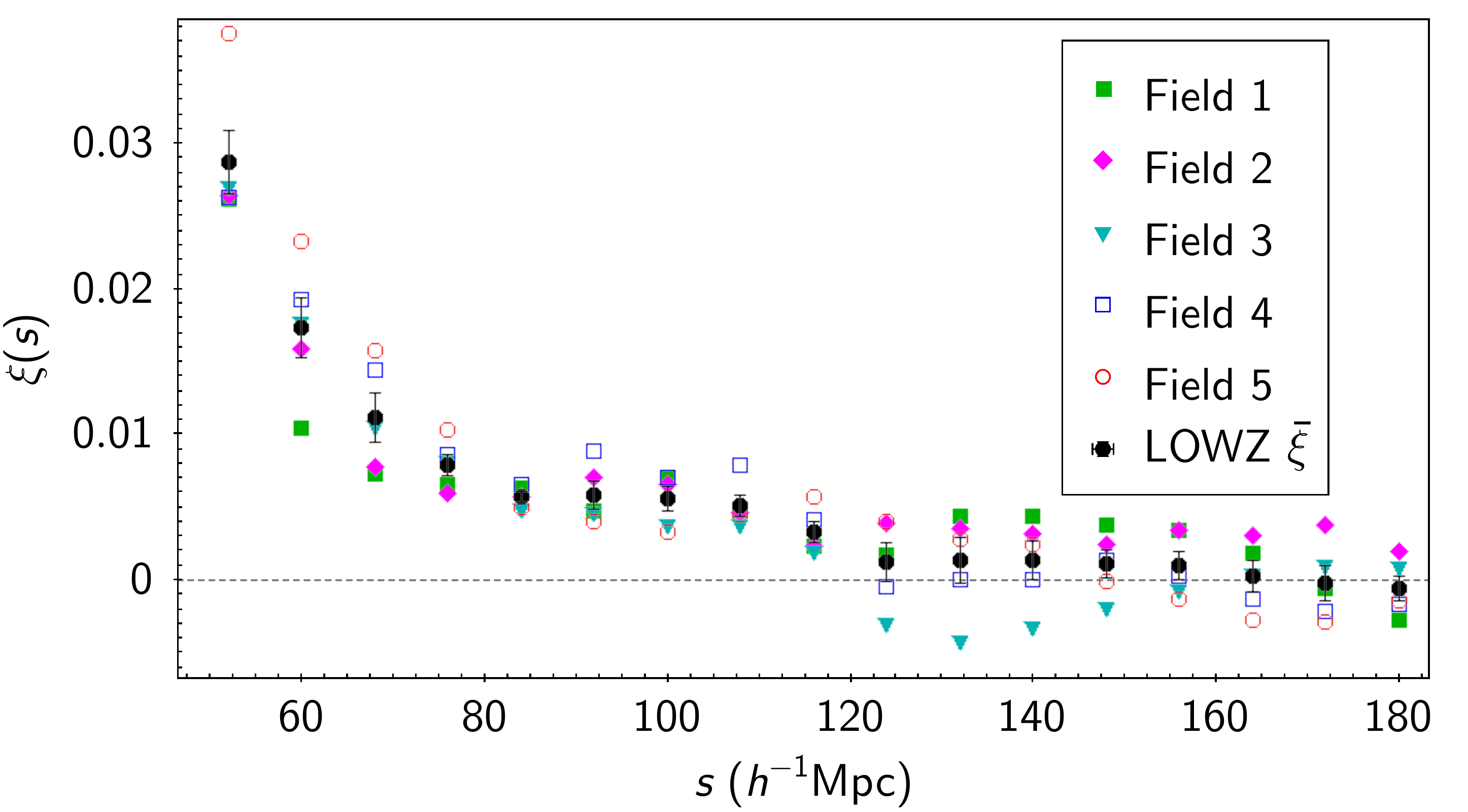}
		\label{fig:sfig_LOWZ_new}
	\end{subfigure}
	\caption{The monopole correlation functions for the individual fields (1: filled squares, 2: filled diamonds, 3: filled triangles, 4: open squares, 5: open circles) and the corresponding mean correlation function (black filled circles), of (a) CMASS and (b) LOWZ samples. The error bars on the mean correlation functions are the standard error on the mean.}
	\label{fig:all_fields_abstract}
\end{figure}

The correlation functions of the individual fields for the LOWZ and CMASS samples along with the corresponding mean correlation functions are displayed in Fig.~\ref{fig:all_fields_abstract}. We note that there is quite a wide variation in these correlation functions with e.g. field 4 for CMASS showing high values at $\sim80h^{-1}$Mpc. In the following sections we compare our measurement of the mean correlation functions with the measurements of  \cite{Cuesta2016}, perform fitting to the mean correlation functions and analyse various aspects of the fitting procedure. Furthermore, we obtain measurements of $D_V(z)$ based on our measured position of the BAO peak. 

\subsection{Comparison with \citet{Cuesta2016}}

\label{sec:Cuesta Comparison}

\begin{figure}
	\begin{subfigure}{\columnwidth}
		\centering
		\caption{}
		\includegraphics[width=\columnwidth]{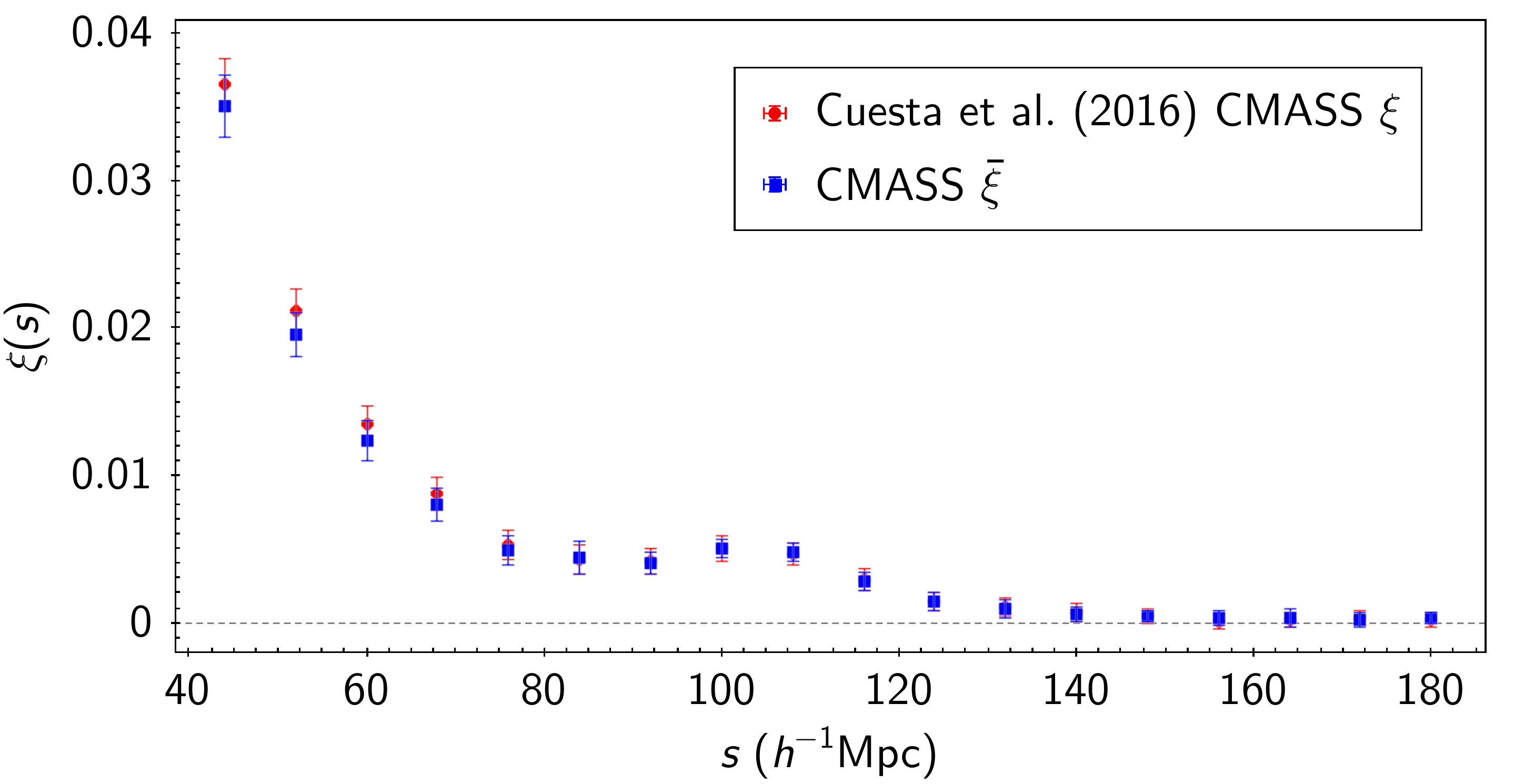}
		\label{fig:CMASS_us_vs_Cuesta.pdf}
	\end{subfigure}	
	\begin{subfigure}{\columnwidth}
		\centering
		\caption{}
		\includegraphics[width=\columnwidth]{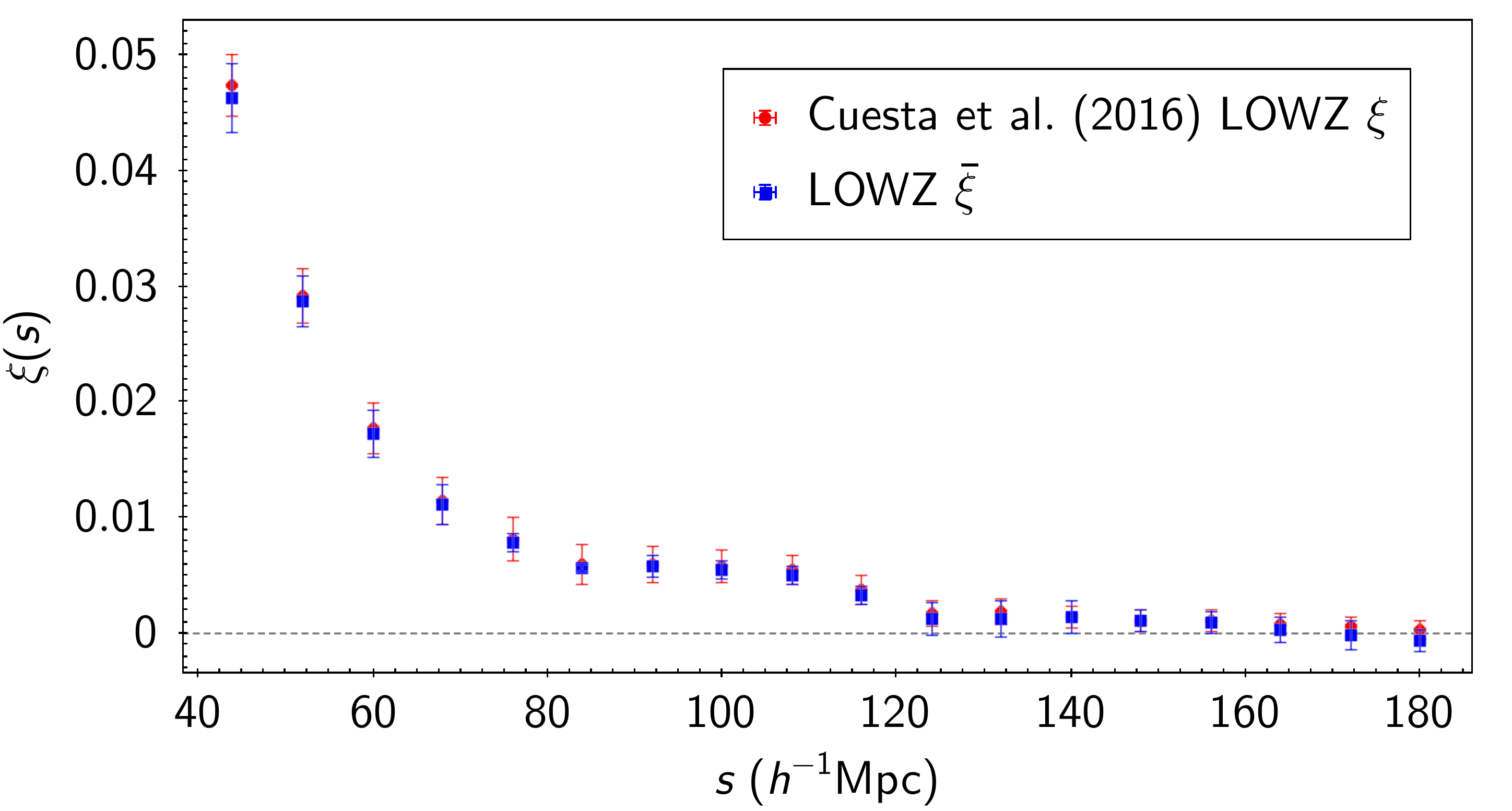}
		\label{fig:LOWZ_us_vs_Cuesta.pdf}
	\end{subfigure}
	\caption[Our measured LOWZ and CMASS $\bar{\xi}(s)$ in comparison with \citet{Cuesta2016}]{A comparison of our mean monopole correlation functions (blue squares) for (a) DR12 CMASS and (b) LOWZ samples, and the pre-reconstruction correlation functions presented in Fig. 1 of \citet{Cuesta2016} (red circles) for these samples. Error bars on our measurements represent the standard error on the mean based on our five subsamples, while the error bars on the \citet{Cuesta2016} data points are based on the BOSS DR12 covariance matrix, obtained from simulated mocks}.
	\label{fig:LOWZ+CMASS_us_vs_Cuesta.pdf}
\end{figure}

Fig.~\ref{fig:LOWZ+CMASS_us_vs_Cuesta.pdf} shows a comparison between our mean correlation functions and the correlation functions obtained by \citet{Cuesta2016} for the DR12 LOWZ and CMASS samples. We find that our measured correlation functions are in excellent agreement with those presented in \citet{Cuesta2016} and we observe no significant changes when we replace the BOSS DR12 randoms with randoms generated by CUTE. Furthermore, we observe no significant variations when we do not apply any weights to the data or randoms. This outcome is however expected due to the high completeness of $98.8\%$ and $97.2\%$ for the CMASS and LOWZ samples respectively \citep[see Fig. 8 of][]{Reid2016}. 


\subsection{Error Analysis Results}

\label{sec:Error Analysis Results}

This section contains a comparison between our two measures of uncertainties (standard error and bootstrap resampling) on the mean correlation functions of the LOWZ and CMASS samples. Here we also include the bootstrap uncertainties based on dividing the CMASS sample into 30 subsamples (see figures in Appendix~\ref{sec:CMASS 30 Fields}). We distinguish between the two bootstrap uncertainties using the labels \lq{CMASS 5}\rq\ and \lq{CMASS 30}\rq.  More importantly comparisons are drawn between our measured empirical errors and errors obtained from simulations presented in \cite{Cuesta2016} for the correlation functions of the LOWZ and CMASS samples. In order to account for the fact that our selected fields do not cover the entire sample area, when comparing our results with those from \cite{Cuesta2016} we scale our measured errors by the square root of the ratio of the total coverage area of our fields to the total sample area.

\begin{figure*}
	\begin{subfigure}{\columnwidth}
		\centering
		\caption{CMASS Uncertainties}
		\includegraphics[width=1\linewidth]{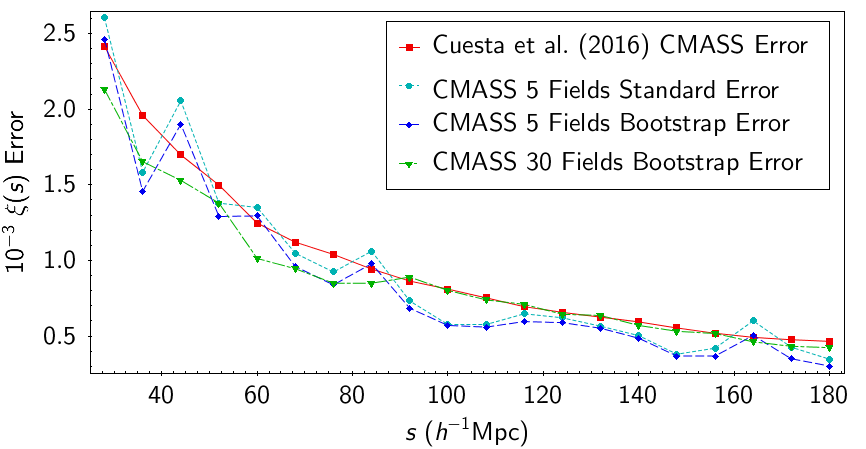}
		\label{fig:CMASS_errors.pdf}
	\end{subfigure}
	\begin{subfigure}{\columnwidth}
		\centering
		\caption{LOWZ Uncertainties}
		\includegraphics[width=1\linewidth]{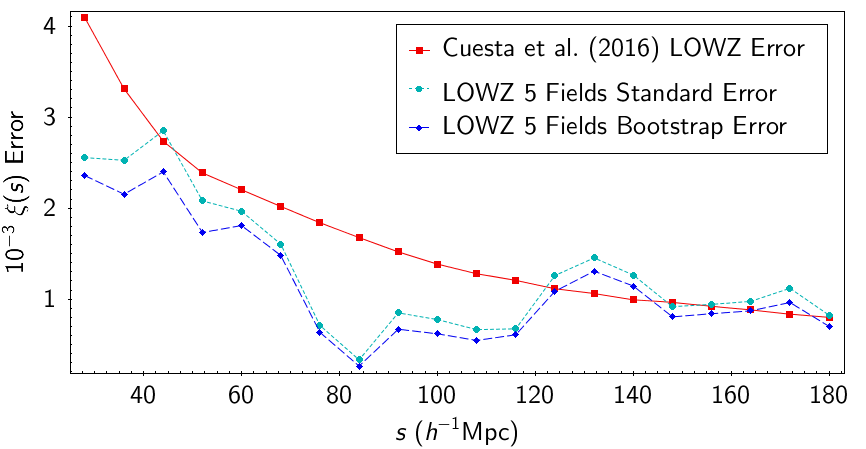}
		\label{fig:LOWZ_errors.pdf}
	\end{subfigure}
	\begin{subfigure}{\columnwidth}
		\centering
		\caption{CMASS Uncertainties Ratio}
		\includegraphics[width=1\linewidth]{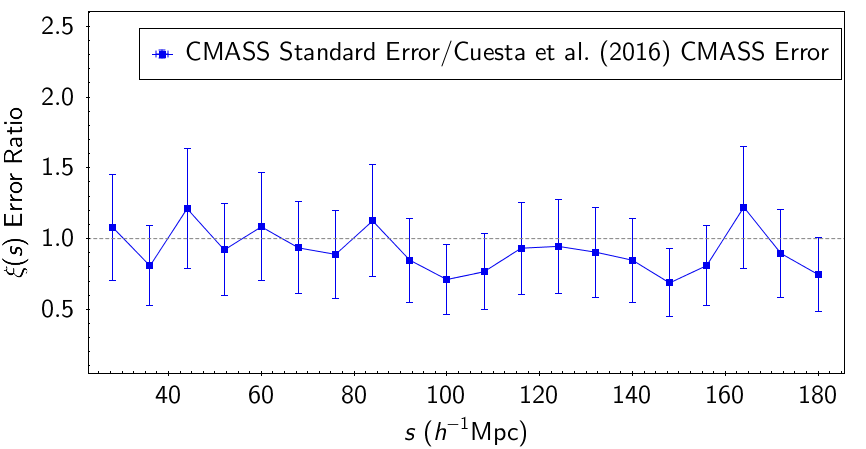}
		\label{fig:CMASS_errors_ratio.pdf}
	\end{subfigure}
	\begin{subfigure}{\columnwidth}
		\centering
		\caption{LOWZ Uncertainties Ratio}
		\includegraphics[width=1\linewidth]{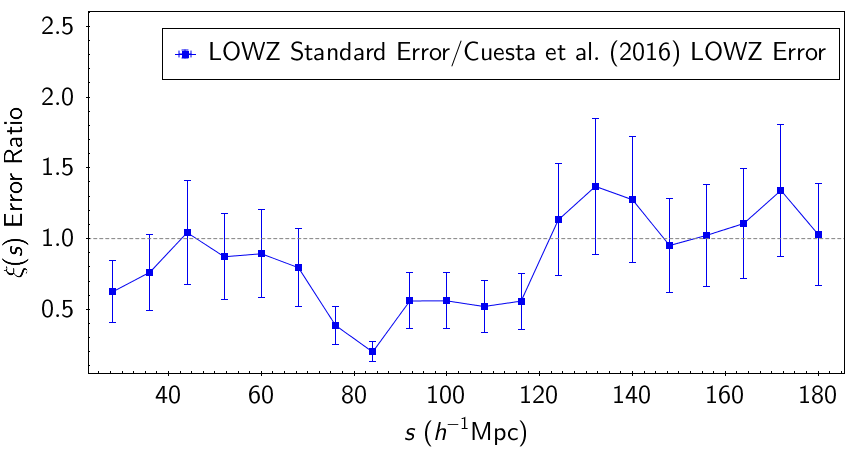}
		\label{fig:LOWZ_errors_ratio.pdf}
	\end{subfigure}
	
	\caption[Comparison of various uncertainty estimates on the CMASS $\bar{\xi}(s)$]{A comparison of the uncertainties on our measured mean correlation function of the CMASS sample, at our primary fitting range  $28\leqslant s \leqslant180h^{-1}$Mpc. The standard error on the mean (light blue circles) and bootstrap (dark blue diamonds) estimates of error for the 5 fields appear to be in good agreement. The bootstrap error from the 30 fields (green inverted triangles) and the uncertainties on the measured correlation functions of \citet{Cuesta2016} (Fig.1) (red squares) are also plotted, showing excellent agreement between the two at scales larger than $90\ h^{-1}$Mpc. Here all our measured errors are scaled by the square root of the ratio of the area covered by our selected fields, to the total sample area (e.g. in the case of 5 fields CMASS by $\sqrt{0.905}$). Subplot (c) shows the ratio of our standard error to the errors presented by \citet{Cuesta2016} for the CMASS sample. Here the error bars represent the error in the error (see the discussion in Section~\ref{sec:Error Analysis Results}). Subplots (b) and (d) contain the equivalent results for the LOWZ sample.}
	\label{fig:errors_ratio.pdf}
\end{figure*}

As shown in Figs.~\ref{fig:CMASS_errors.pdf} and ~\ref{fig:LOWZ_errors.pdf}, we find a good agreement between the standard error and bootstrap error estimates for both samples. Fig.~\ref{fig:CMASS_errors.pdf} shows that at our main scale of interest (in the vicinity of the $108\ h^{-1}$Mpc bin where the BAO peak lies), our results for the 5 fields CMASS sample also appear to be in reasonable agreement with the errors presented by \cite{Cuesta2016}. Furthermore, our 30 fields bootstrap uncertainties appear to be in excellent agreement with those from \cite{Cuesta2016} at scales larger than $90\ h^{-1}$Mpc. To provide a quantitative demonstration of the level agreement between the errors from \cite{Cuesta2016} and the simple case of standard errors obtained from 5 fields, we make use of the fractional error in the error, given by $1/\sqrt{2N-2}$ \citep{Squires}. Here $N$ is the number of measurements (in our case 5), giving a fractional error in the error of $\approx35\%$. Fig.~\ref{fig:CMASS_errors_ratio.pdf} shows the ratio of our measured standard error to the errors presented by \cite{Cuesta2016} for the CMASS sample with the error bars being the error on our measured standard error. We can see that at the $108\ h^{-1}$Mpc bin this ratio is $0.8$ which is consistent with unity within the error bars, and the general agreement between the errors is an indication that the QPM mocks reproduce an accurate representation of the data. As shown in Fig.~\ref{fig:LOWZ_errors_ratio.pdf} however, in the case of the LOWZ sample the ratio between the two errors varies to a greater extent as a function of scale, with the discrepancy between the two errors being larger around the BAO scale. This indicates that the errors presented by \cite{Cuesta2016} do not appear to be underestimated in this region.

\subsection{Data Fitting Results}
\label{sec:Data fitting results}

The best-fit values of $\alpha$ obtained from fitting the data with various models, across 20 bins, with centres in the range $28\leqslant s \leqslant180h^{-1}$Mpc are summarised in Table~\ref{tab:alphas_table}\footnote{Note that we place the main focus of our analysis on the results corresponding to this fitting range in order to match the fitting range chosen in \citet{Cuesta2016}, allowing for direct comparison of the results. As discussed in Section~\ref{sec:Fitting the Correlation Function}, when fitting the correlation functions we use the BOSS DR12 covariance matrix used in the analysis of \cite{Cuesta2016}.}. The pre-reconstruction best fit values of $\alpha$ from \citet{Cuesta2016} are included in this table for comparison. Here, \lq $\alpha$\rq\ refers to values obtained from fitting to the mean correlation functions of the LOWZ and CMASS samples, with errors given by the procedure described in Section~\ref{sec:Fitting the Correlation Function}. The \lq 5-fields $\bar{\alpha}$\rq\ values in this table are obtained by fitting to the correlation functions of each field individually resulting in 5 measurements of $\alpha$ (these are presented in Table~\ref{tab:5_filds_alphas_table}), and calculating the mean and standard error of these measurements. When fitting to correlation functions of individual fields we scale the BOSS DR12 covariance matrix by a factor of 5. 

As shown in Table~\ref{tab:alphas_table}, we find that our measured \lq 5-fields $\bar{\alpha}$\rq\ values are in good agreement with our overall values of $\alpha$. This demonstrates the robustness of the implemented fitting procedure in producing an accurate measurement of the position of the BAO peak. Furthermore, when comparing the results corresponding to fits with the complete model, we find that for both CMASS and LOWZ samples, our measured values of $\alpha$ are in agreement with the measurement presented by \cite{Cuesta2016}, with the errors on $\alpha$ being similar in size.

\begin{table*}
	\centering
	\caption[Our measurements of the BAO peak position ($\alpha$) in comparison to \citet{Cuesta2016} ]{Results of fitting the correlation functions of the LOWZ and CMASS samples using the complete $\xi^{fit}$ model described in Eq.~\ref{eq:xi_fit} and the same model without the $A(s)$ nuisance fitting parameters. In line with \citet{Cuesta2016} the fitting is performed in the range $28\leqslant s \leqslant180h^{-1}$Mpc. Here the \lq $\alpha$\rq\ values are obtained from fitting to the mean correlation function $\bar{\xi}(s)$ of each sample, presenting the corresponding $\chi^2_{min}$ over the number of degrees of freedom and \lq Significance\rq\ refers to the significance of the detection of the BAO peak, using the complete fitting model (see Section~\ref{sec:Significance of BAO Peak Detection}). The $F$-ratio $p$-values (given by Eq.~\ref{eq:F_ratio}) indicate the probability that the nuisance parameters do not contribute to the goodness-of-fit of the full model. The \lq 5-fields $\bar{\alpha}$\rq\ values are based on taking the mean and standard error of the individual $\alpha$s, measured from fits to correlation functions of the 5 fields in the LOWZ and CMASS samples (see Table~\ref{tab:5_filds_alphas_table}). We have used the BOSS DR12 covariance matrices in our fits scaling them by a factor of 5 when fitting to the 5 fields individually. For comparison, the best-fit values of $\alpha$ from \citet{Cuesta2016} (Table 10), for the pre-reconstruction LOWZ and CMASS sample are also included.}
	\label{tab:alphas_table}
	\begin{tabular}{ll|cccc|c} 
		\hline
		\hline
		This Work &  Model & $\alpha$ & $\chi^2_{min}/dof$ & Significance & $F$-ratio & 5-fields $\bar{\alpha}$ \\
		\hline
		CMASS & $B^2\xi_m+A(s)$ & $1.0109\pm0.0121$ & 14.9/15 & $8.0\sigma$ & $4.56\ (p=0.018)$ & $1.0122\pm0.0172$\\
		& $B^2\xi_m$ & $1.0009\pm0.0116$ & 28.5/18 & $6.9\sigma$ & & $1.0021\pm0.0101$\\
		\hline
		LOWZ & $B^2\xi_m+A(s)$ & $1.0074\pm0.0266$ & 15.5/15 & $4.3\sigma$ & $9.68\ (p=0.00084)$ & $1.0050\pm0.0421$\\
		& $B^2\xi_m$ & $0.9698\pm0.0523$ & 45.5/18 & $1.8\sigma$ & & $1.0060\pm0.0195$\\
		\hline
		\hline
		Cuesta et al. (2016)  &  Model & $\alpha$ & $\chi^2_{min}/dof$ & Significance &&  \\
		\hline
		CMASS & $B^2\xi_m+A(s)$ & $1.0153\pm0.0134$ & 12/15 & $8.0\sigma$ && \\
		LOWZ & $B^2\xi_m+A(s)$ & $1.0085\pm0.0300$  & 13/15 & $4.0\sigma$ &&\\
	\end{tabular}
\end{table*}
  
\begin{table*}
	\centering
	\caption[Results of fitting the ${\xi}(s)$ of the individual fields in the CMASS and LOWZ samples]{Results of fitting the correlation functions of the 5 individual fields in the LOWZ and CMASS samples using two different models, over the range $28\leqslant s \leqslant180h^{-1}$Mpc. Here we have used the BOSS DR12 covariance matrices, scaled by a factor of 5 and \lq Significance\rq\ refers to the significance of the detection of the BAO peak, using the complete fitting model (see Section~\ref{sec:Significance of BAO Peak Detection}). The mean $\alpha$ and its standard error obtained based on the values of $\alpha$ in this table are presented under the \lq 5-fields $\bar{\alpha}$ \rq\ column in Table~\ref{tab:alphas_table}.}
	\label{tab:5_filds_alphas_table}
	\begin{tabular}{c|clccc} 
		\hline
		\hline
		This Work & Field & Model & $\alpha$ & $\chi^2_{min}/dof$& Significance \\
		\hline
		& 1 & $B^2\xi_m+A(s)$ & $1.0070\pm0.0207$ & 14.5/15 & $4.2\ \sigma$  \\
		&  & $B^2\xi_m$ & $1.0034\pm0.0219$ & 21.7/18 &   \\
		
		& 2 & $B^2\xi_m+A(s)$ & $0.9656\pm0.0245$ & 16.2/15 & $3.3\ \sigma$ \\
		&  & $B^2\xi_m$ & $0.9751\pm0.0279$ & 21.1/18 &  \\
		
		CMASS & 3 & $B^2\xi_m+A(s)$ & $0.9924\pm0.0406$ & 12.6/15 & $2.9\ \sigma$ \\
		&  & $B^2\xi_m$ & $0.9848\pm0.0273$ & 12.2/18 &   \\
		
		& 4 & $B^2\xi_m+A(s)$ & $1.0703\pm0.0506$ & 13.5/15 & $2.0\ \sigma$ \\
		&  & $B^2\xi_m$ & $1.0273\pm0.0380$ & 29.7/18 &   \\
		
		& 5 & $B^2\xi_m+A(s)$ & $1.0319\pm0.0258$ & 10.8/15 & $3.3\ \sigma$ \\
		&  & $B^2\xi_m$ & $1.0221\pm0.0261$ & 12.2/18 &  \\
		\hline
		& 1 & $B^2\xi_m+A(s)$ & $1.0535\pm0.0703$ & 34.6/15 & $1.3\ \sigma$  \\
		&  & $B^2\xi_m$ & $1.0403\pm0.0711$ & 35.3/18 &   \\
		
		& 2 & $B^2\xi_m+A(s)$ & $1.1049\pm0.1159$ & 23.7/15 & $1.6\ \sigma$  \\
		&  & $B^2\xi_m$ & $1.0624\pm0.0290$ & 33.8/18 &  \\
		
		LOWZ & 3 & $B^2\xi_m+A(s)$ & $1.0117\pm0.0765$ & 17.0/15 & $1.8\ \sigma$ \\
		&  & $B^2\xi_m$ & $0.9817\pm0.0344$ & 16.3/18 &  \\
		
		& 4 & $B^2\xi_m+A(s)$ & $1.0210\pm0.0361$ & 22.3/15 & $2.3\ \sigma$ \\
		&  & $B^2\xi_m$ & $1.0009\pm0.0363$ & 26.0/18 &  \\
		
		& 5 & $B^2\xi_m+A(s)$ & $0.8674\pm0.0934$ & 18.2/15 & $1.7\ \sigma$ \\
		&  & $B^2\xi_m$ & $0.9527\pm0.0125$ & 43.5/18 &  \\
		
	\end{tabular}
\end{table*}

We find the values of $\alpha$ measured for the individual fields in Table~\ref{tab:5_filds_alphas_table} to be in general agreement with the measurements of $\alpha$ from \citealt{Cuesta2016}. In cases where there appears to be a divergence between the measurements, (for instance our result of fitting the correlation function of field 4 in the CMASS sample with the complete model appears to be $\approx 1.7\sigma$ away from the value of $\alpha$ measured by \citealt{Cuesta2016}), the dependency seems to be due to the shape of the BAO peak (which in this case appears to be relatively flat, as seen in Fig.~\ref{fig:sfig_CMASS_new}). However, as the \lq 5-fields $\bar{\alpha}$\rq\ values are in agreement with the measurements of $\alpha$ from the mean correlation functions, these effects seem to cancel out when we take the average over the 5 fields, even given our relatively small number of subsamples. 

The performance of the two models in fitting the correlation functions (given by the $\chi^2_{min}/dof$ goodness of fit indicator) also appear to vary largely depending on the shape of the correlation function. However, with the exception of certain fields (for instance field 3 of both CMASS and LOWZ samples), the complete model appears to perform better overall in providing good fits. It is important to note however, that the performance of a model in providing a good fit is not necessarily indicative that the correlation function has provided a representative and accurate measurement of $\alpha$, and one should also consider the shape and prominence of the BAO peak in the correlation function itself\footnote{In Section~\ref{sec:Significance of BAO Peak Detection} we discuss how the shape of the $\Delta\chi^2$ curve could also provide a measure of the degree to which we could be confident in our measurement of $\alpha$.}. This is exemplified by field 4 in the CMASS sample where the $\chi^2_{min}/dof$ value indicates that the complete model has provided a reasonably good fit to the data but due to the shape of the correlation function (see Fig.~\ref{fig:sfig_CMASS_new}), an accurate determination of the position of the peak has not been possible. Finally we find that the significance of detection of the peak in the individual fields to be generally lower than the significance of the detection of the peaks in the mean correlation functions of the two samples (as shown in Table~\ref{tab:alphas_table}). This is a further indication of the lack of prominent and well defined peaks in the correlation functions of the individual fields and as shown once again by field 4 in the CMASS sample, a low significance of detection of the peak could also hint towards the potential unreliability of the measured $\alpha$.

\subsection{Model Comparison}
\label{sec:Model Comparison}

\begin{figure}
	\begin{subfigure}{\columnwidth}
		\centering
		\caption{}
		\begin{tikzpicture}
		\node[anchor=south west,inner sep=0] (image) at (0,0) {\includegraphics[width=0.98 \textwidth]{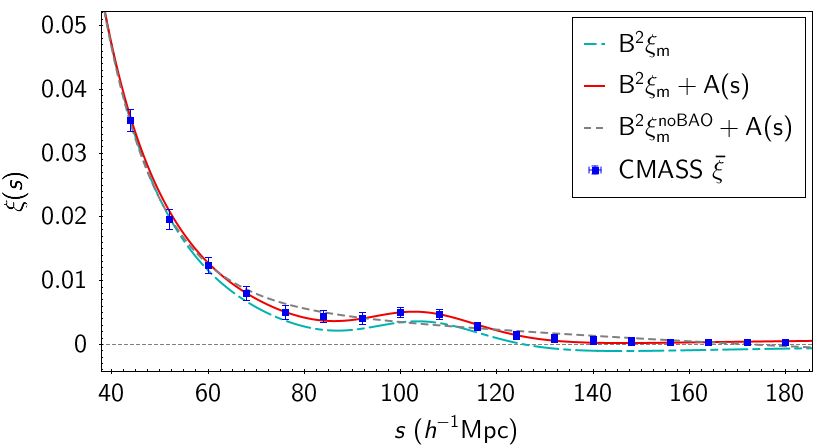}};
		\begin{scope}[x={(image.south east)},y={(image.north west)}]
		\node[anchor=south west,inner sep=0] (image) at (0.210,0.444) {\includegraphics[width=0.48\textwidth]{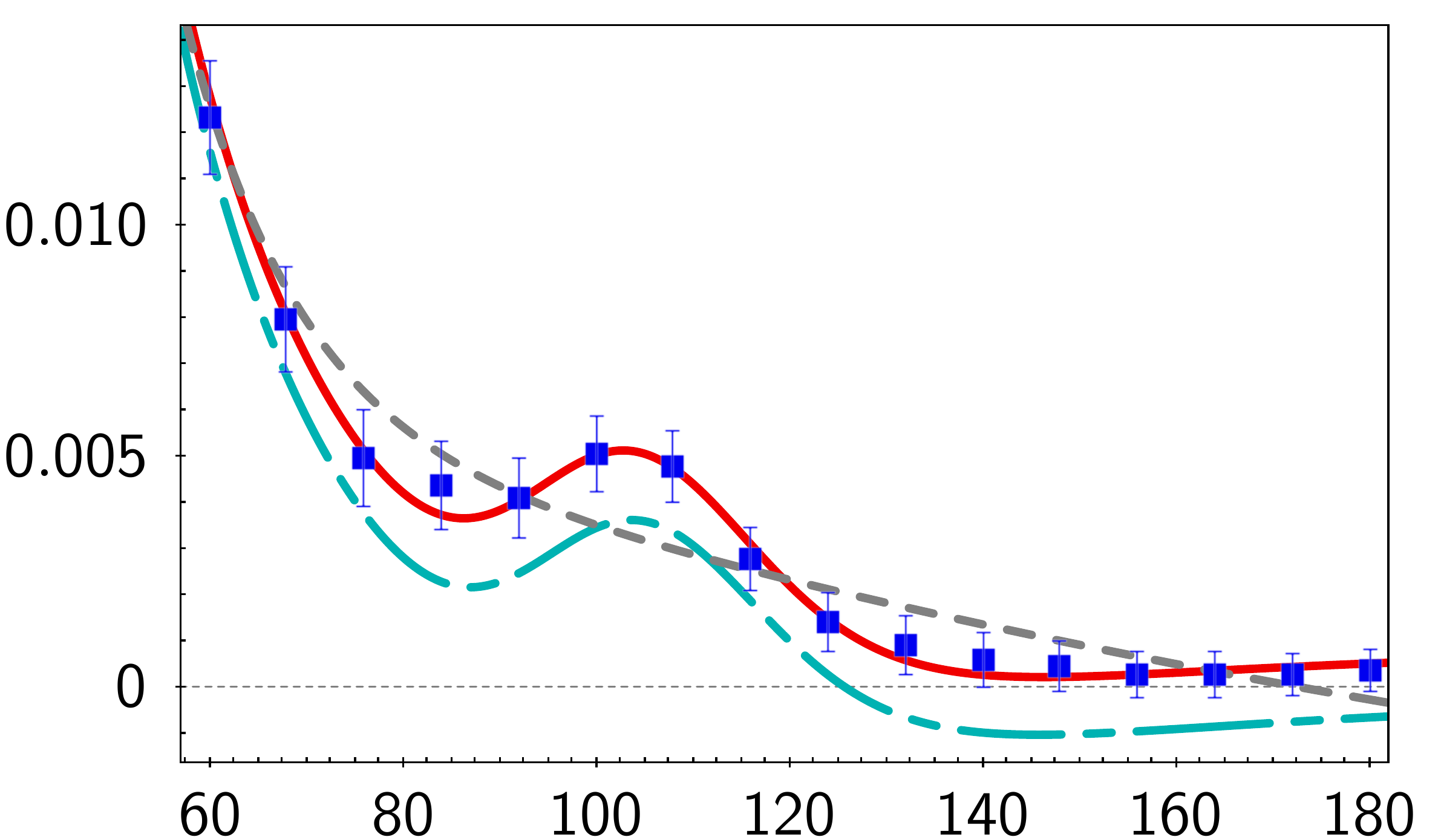}};
		\end{scope}
		\end{tikzpicture}\\ 
		\label{fig:CMASS_fits.pdf}
	\end{subfigure}
	\begin{subfigure}{\columnwidth}
		\centering
		\caption{}
		\begin{tikzpicture}
		\node[anchor=south west,inner sep=0] (image) at (0,0) {\includegraphics[width=\textwidth]{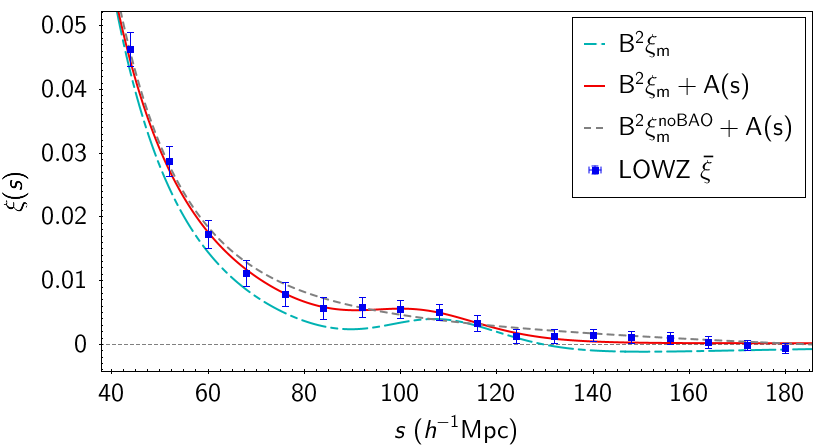}};
		\begin{scope}[x={(image.south east)},y={(image.north west)}]
		\node[anchor=south west,inner sep=0] (image) at (0.227,0.464) {\includegraphics[width=0.47\textwidth]{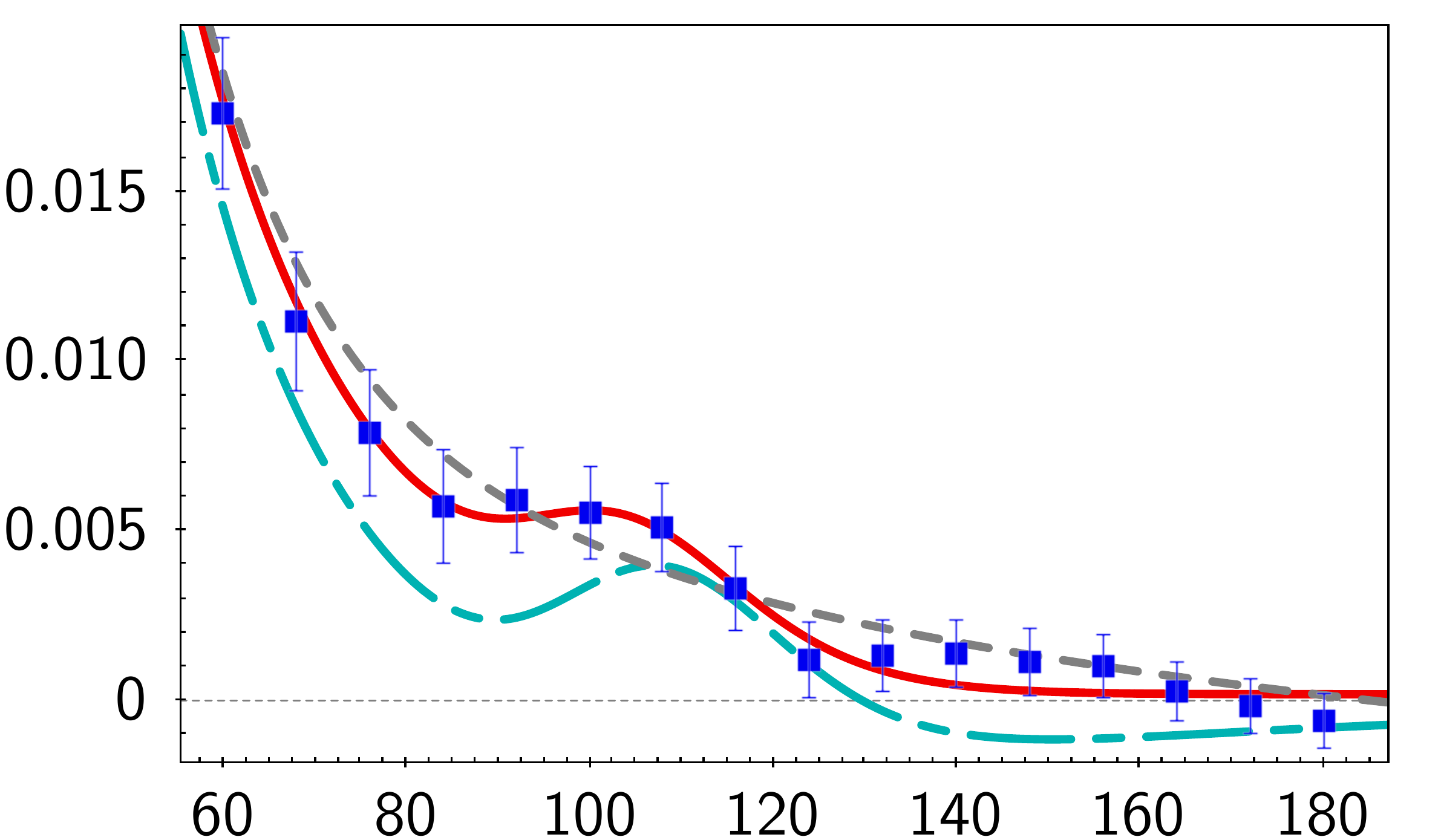}};
		\end{scope}
		\end{tikzpicture}\\
		\label{fig:LOWZ_fits.pdf}
	\end{subfigure}
	\caption[Fitting the LOWZ and CMASS $\bar{\xi}(s)$ with various models]{The results of fitting the mean correlation function of (a) CMASS and (b) LOWZ samples with various fitting models in the range $28\leqslant s \leqslant180h^{-1}$Mpc. The blue dot-dashed curve is the $\xi^{fit}$ model (equation \ref{eq:xi_fit}) with the $B$ fitting parameter only, while the red solid curve shows the same model fitted with both $B$ and $A(s)$ fitting terms. The grey dashed curve is the $\xi^{noBAO}$ model fitted with the $B$ and $A(s)$ fitting terms. The error bars shown are the square root of the diagonal elements of the BOSS DR12 covariance matrices.} 
	\label{fig:CMASS_60-124_model_comparison.pdf}
\end{figure}

Fig.~\ref{fig:CMASS_60-124_model_comparison.pdf} shows the results of fitting the mean correlation functions of the CMASS and LOWZ samples with the $\xi^{fit}$ model, fitted with and without the $A(s)$ nuisance parameters, and the $\xi^{noBAO}$ model fitted with both $B$ and $A(s)$ fitting terms. The important role played by the $A(s)$ nuisance fitting terms in producing a good fit is highlighted in these plots. This is also demonstrated numerically in Table~\ref{tab:alphas_table}, with the fits without the $A(s)$ having increased $\chi^2_{min}/dof$ values indicating the lower quality of fits. We assess the $\chi^2_{min}/dof$ statistic based on the corresponding $p$-value = $1-p(\chi^2(dof)\geq\chi^2_{min}|H)$, which is defined as the probability of obtaining a $\chi^2(dof)$ value at least as extreme as the value obtained, given our null hypothesis $H$: that the data is consistent with the model. In other words, the $p$-value is the probability of obtaining the observed data, under the assumption that the model is correct, and a measure of the significance at which the model is rejected by the data is given by $1-p$-value.

We note that the visual impression given in Fig.~\ref{fig:CMASS_fits.pdf} is that the $\Lambda$CDM model without nuisance parameters for the CMASS sample is rejected at a higher significance than by the $28.5/18$ ($p=0.055$) indicated in Table~\ref{tab:alphas_table}. Indeed, when only the diagonal terms of the covariance matrix are used in the fitting, the significance of rejection rises to $64.9/18$ ($p=3.23\times10^{-7}$) (see Table~\ref{tab:covmat}). Thus in this case the inclusion of the full covariance matrix causes a large reduction in $\chi^2_{min}/dof$.

\begin{table*}
	\centering
	\caption[]{BAO peak detection significance using various models, along with corresponding $\chi^2_{min}/dof$ and $p$-values obtained from fitting the mean correlation functions of the CMASS and LOWZ samples in the range $28\leqslant s \leqslant180h^{-1}$Mpc, using the full covariance matrix and the diagonal elements of the matrix only. As fitting with the full covariance matrix and the diagonal elements only could result in different best fit models, in order to ensure the fairness of the comparison, when calculating the 'Diagonal Elements' $\chi^2_{min}/dof$ values, we use  best fit models obtained using the full covariance matrices. Furthermore, we quote the $\chi^2_{min}/dof$ values at fixed values of $\alpha$ corresponding to our measurements of the BAO peaks from Table~\ref{tab:alphas_table} (e.g. in the case of CMASS, at 0.9892 and 0.9991 for our full and reduced models respectively).}
	\label{tab:covmat}
	\begin{tabular}{ll||ccc|ccc} 
			\hline
			\hline
			& & Full Matrix & & & Diagonal Elements & &  \\
			Sample & Model & $\chi^2_{min}/dof$ & $p$-value & Significance & $\chi^2_{min}/dof$ & $p$-value & Significance  \\
			\hline
			CMASS & $B^2\xi_m+A(s)$ & $14.9/15$ & $4.59\times10^{-1}$ & $8.0\sigma$ & $2.9/15$ & $9.99\times10^{-1}$ & $4.7\sigma$ \\
			& $B^2\xi_m^{noBAO}+A(s)$ & $80.0/15$ & $6.98\times10^{-11}$ & & $25.3/15$ & $4.61\times10^{-2}$ &  \\
			& $B^2\xi_m$& $28.5/18$ & $5.48\times10^{-2}$ & $6.9\sigma$ & $64.9/18$ & $3.23\times10^{-7}$ & $5.8\sigma$ \\
			& $B^2\xi_m^{noBAO}$ & $76.7/18$ & $3.22\times10^{-9}$ & &  $98.8/18$ & $3.67\times10^{-13}$ & \\
			\hline
			LOWZ & $B^2\xi_m+A(s)$ & $15.5/15$ & $4.16\times10^{-1}$ & $4.3\sigma$ & $5.4/15$ & $9.88\times10^{-1}$ & $2.8\sigma$  \\
			& $B^2\xi_m^{noBAO}+A(s)$ & $33.9/15$ & $3.52\times10^{-3}$ & & $13.2/15$ & $5.87\times10^{-1}$ &  \\
			& $B^2\xi_m$ & $45.5/18$ & $3.51\times10^{-4}$ & $1.8\sigma$ & $47.7/18$ & $1.67\times10^{-4}$ & $2.2\sigma$ \\
			& $B^2\xi_m^{noBAO}$ & $48.8/18$ & $1.14\times10^{-4}$ & & $52.8/18$ & $2.82\times10^{-5}$ & \\
		\end{tabular}
\end{table*}

We then take a more detailed look at how significant the nuisance parameters are in achieving a good fit for the $\Lambda$CDM model. Given our two nested fit models, we can make use of the $F$-ratio (see e.g.~\citealt{Gregory2005}) in order to determine whether the use of the more complex model results in a statistically significant improvement in fit quality. The $F$-ratio is given by

\begin{equation}
F= \frac{(\chi^2_{simple}-\chi^2_{complex})/(dof_{simple}-dof_{complex})}{\chi^2_{complex}/dof_{complex}}.
\label{eq:F_ratio}
\end{equation}
Here $\chi^2_{simple}$ and $\chi^2_{complex}$ refer to the $\chi^2_{min}$ values obtained from fitting the $\xi^{fit}$ model without the $A(s)$ nuisance fitting terms, and by the complete $\xi^{fit}$ model respectively, and $dof$ are the degrees of freedom associated with each model. Once the $F$ value is obtained we can test the validity of our null hypothesis that the complex model does not provide a significantly better fit than the simple model. Similar to the $\chi^2$ analysis above, we assess the validity of the null hypothesis based on the $p$-value associated with the resulting $F$ statistic.

Based on the $\chi^2_{min}/dof$ values presented in Table~\ref{tab:alphas_table}, for the fitting range $28\leqslant s \leqslant180 h^{-1}$Mpc, we obtain $F$ values of $4.56\ (p=0.018)$ and $9.68\ (p=0.00084)$ for the CMASS and LOWZ samples respectively. In other words our simple model is rejected in favour of the full $\xi^{fit}$ model by the data, (given that assuming the null hypothesis is correct, i.e. that there is no significant difference between the two models, the probability of obtaining an $F$ statistic at least as extreme as the values here by chance are $\approx 1.8\%$ and $0.1\%$ for the CMASS and LOWZ samples respectively). This means that the inclusion of the nuisance parameters results in a significant improvement to the fit. This is specially true in the case of the LOWZ sample, where as seen in Fig.~\ref{fig:LOWZ_fits.pdf}, the BAO peak in the correlation function appears flatter in the $\approx 80-100 h^{-1}$Mpc range, explaining the strong need for the nuisance parameters at the level of significance indicated by the $F$ test. 

\subsection{Significance of BAO Peak Detection}
\label{sec:Significance of BAO Peak Detection}

\begin{figure*}
	\begin{subfigure}{\columnwidth}
		\centering
		\caption{Full Covariance Matrix}
		\includegraphics[width=1\linewidth]{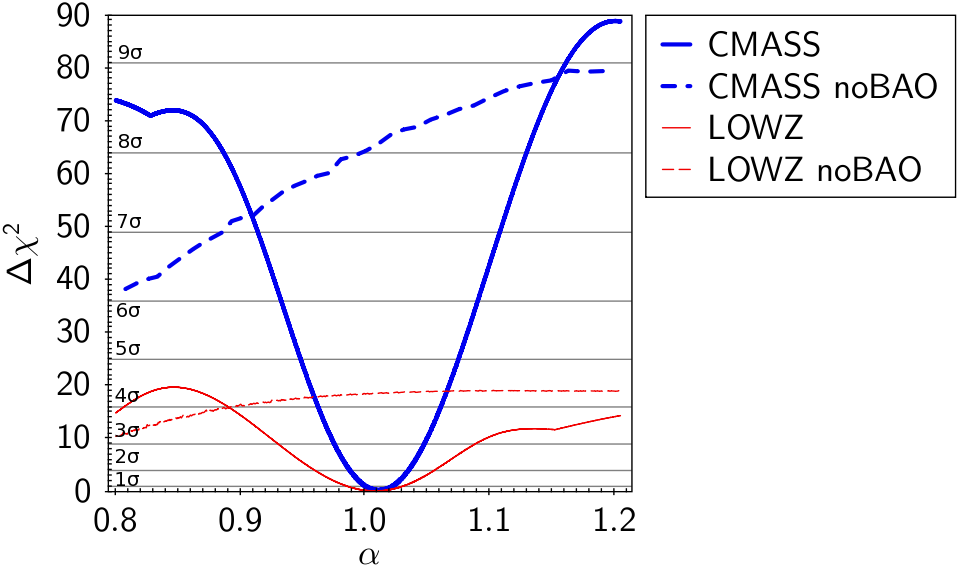}
		\label{fig:LOWZ+CMASS+noBAO_delta_Chi2_02to2.eps}
	\end{subfigure}
	\begin{subfigure}{\columnwidth}
		\centering
		\caption{Diagonal Covariance Matrix Elements}
		\includegraphics[width=1\linewidth]{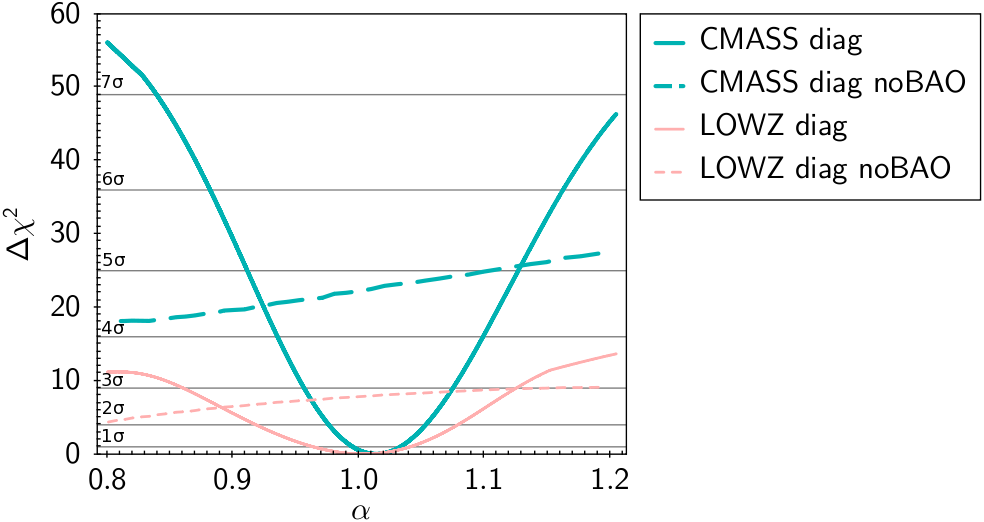}
		\label{fig:LOWZ+CMASS+noBAO_delta_Chi2_08to12_v3_diag.png}
	\end{subfigure}
	\caption[]{(a) Significance of the detection of the BAO feature based on fitting the $\bar{\xi}$ for the LOWZ (red curves) and CMASS (bold blue curves) samples in the range $28\leqslant s \leqslant180h^{-1}$Mpc. The solid lines correspond to fits to the data based on the $\xi^{fit}$ model which contains BAO, while the dashed lines correspond to fits based on the $\xi^{noBAO}$ model with no BAO feature. In all cases the complete models including the $A(s)$ fitting terms and the full covariance matrix are used. Here $\Delta\chi^2=\chi^2(\alpha)-\chi^2_{min}$, where $\chi^2_{min}$ is the minimum $\chi^2$ value using the model containing BAO. Comparing the dashed and solid lines provides a measure of our level of confidence that the BAO feature exists in the data. Here the BAO peak is detected at $\approx4.3\sigma$ for the LOWZ sample and $\approx8\sigma$ for the CMASS sample. (b) Same as (a) but fitting with the diagonal covariance matrix elements only.} 
	\label{fig:delta_chi2} 
\end{figure*}

The $\Delta\chi^2$ curves based on fitting the mean correlation
functions of the CMASS and LOWZ samples, with the $\xi^{fit}$ and
$\xi^{noBAO}$ models are presented in Fig.~\ref{fig:LOWZ+CMASS+noBAO_delta_Chi2_02to2.eps}. Here the complete fitting models including the $A(s)$ fitting terms are used and $\Delta\chi^2=\chi^2(\alpha)-\chi^2_{min}$, where $\chi^2_{min}$ is the minimum $\chi^2$ value using the model containing BAO. A comparison of the two models shows that we detect the BAO peak in the data at an $\approx4.3\sigma$ level for the LOWZ sample and at $\approx8\sigma$ for the CMASS sample, in agreement with the findings of \cite{Cuesta2016}. Note that we measure the BAO peak detection significance at the best-fit value of $\alpha$ given by the model containing BAO. A second test of BAO significance is also captured in Fig.~\ref{fig:LOWZ+CMASS+noBAO_delta_Chi2_02to2.eps}. For the CMASS sample, it can be seen from the plateau height of the $\Delta\chi^2$ curve (solid blue line), that local maximum lies at a value of $\approx72$ above the minimum, meaning that we can apparently be confident in our measured best-fit value of $\alpha$ at
$\approx8.5\sigma$. For the LOWZ sample, the maximum lies at
$\approx20$, indicating that our best fit value of $\alpha$ is preferred at $\approx4.0\sigma$ by the data. These values are usually taken to indicate that we have obtained well-constrained measurements of $\alpha$ in both cases. In the case of the LOWZ sample it can also be seen that the plateau is lower on the left hand side ($\alpha<0.9$) in comparison to the plateau on the right hand side ($\alpha>1.1$). This is once again a consequence of the flatness of the BAO peak in the LOWZ correlation function in the scales of $80<s<100h^{-1}$Mpc, as discussed in the previous section.

However, the level of scatter between field-to-field correlation functions around the BAO peak (as shown in Fig.~\ref{fig:all_fields_abstract}), prompts us to caution that $8\sigma$ and $4\sigma$ BAO peak significances for CMASS and LOWZ may be over-optimistic. Although, it must be remembered that these significances are calculated after the fitting of nuisance parameters  which will clearly remove long-wavelength artefacts that otherwise can add to the noisy impression given by individual fields in Fig.~\ref{fig:all_fields_abstract}.

Another consideration might also involve the anomalously low $\chi^2_{min}/dof=2.9/15$ recorded for our best fit to the CMASS sample with our fiducial plus nuisance parameters model, using the diagonal covariance matrix elements only, compared to $14.9/15$ using the full matrix, as shown in Table~\ref{tab:covmat} (with the associated $\Delta\chi^2$ plot shown in Fig.~\ref{fig:LOWZ+CMASS+noBAO_delta_Chi2_08to12_v3_diag.png}). In the case of using the diagonal elements, the full noBAO model is also only rejected at $\chi^2_{min}/dof=25.3/15$ ($p\approx0.046$), in comparison to the much higher rejection using the full covariance matrix $\chi^2_{min}/dof=80.0/15$ ($p\approx6.98\times10^{-11}$).

We find similar results for the LOWZ sample with a reduction from
$\chi^2_{min}/dof=15.5/15$ to $\chi^2_{min}/dof=5.4/15$ for our full
model using the full and diagonal matrices respectively. We also record a notable reduction in the level of rejection of the noBAO model by the LOWZ data, from $\chi^2_{min}/dof=33.9/15$ ($p\approx3.5\times10^{-3}$) using the full matrix, to a good fit with $\chi^2_{min}/dof=13.2/15$ ($p\approx0.59$) using the diagonal elements.
 
For models with nuisance fitting parameters, using the full covariance matrix appears to increase the $\chi^2_{min}/dof$ significantly compared to using the diagonal terms only. This is opposite to what is seen in other cases such as the fit of the fiducial $\Lambda$CDM model where no nuisance parameters are used (see Table~\ref{tab:covmat}). This may be due to having both positive and negative fit residuals in the first case and residuals mainly of one sign in the latter case, and a covariance matrix with exclusively positive elements. Given the size of this effect, we perform a further test by replacing the off-diagonal CMASS covariance matrix elements by zero, increasingly far from the diagonal (leaving a `band' matrix). This is motivated by the correlation matrix in Fig.~\ref{fig:Correlation_matrix.png} showing that the covariance elements decrease systematically away from the diagonal. We found that $\chi^2_{min}\approx3$ maintained when up to the first 14 off-diagonal elements were retained and only increased to $\chi^2_{min}\approx15$ when elements 15-20 (indicated in Fig.~\ref{fig:Correlation_matrix.png} by the red outline) were included. This effect also appears to be important for assigning the significance of BAO peak detection (as shown in Table~\ref{tab:covmat}, reducing the detection significance from $8.0\sigma$ to $4.7\sigma$ and from $4.3\sigma$ to $2.8\sigma$ for the CMASS and LOWZ when using the complete fitting model). For the CMASS sample, we observe a similar jump in the significance of peak detection from $\approx3.5\sigma$ when only the first 13 off-diagonal elements were included, to $\approx8\sigma$ once elements 14 and higher are included. We note that here the main contribution to the increase in $\Delta\chi^2$ (and hence the peak detection significance) appears to be from the large increase in the  $\chi^2_{min}$ of the noBAO model which rises by $\sim70$, while the $\chi^2_{min}$ of the model containing BAO only rises by 6.5. One can similarly see this in Table~\ref{tab:covmat} with the large increase of $\sim55$ in the $\chi^2_{min}$ of the noBAO+$A(s)$ model compared to only $12$ for the BAO+$A(s)$ model, as we go from fitting with the diagonal elements only to using the full matrix. The sharp nature of this increase and its marked effect on the significance of model rejection may seem somewhat anomalous, given that one would expect relatively low correlation between $\xi(s)$ points $\approx100h^{-1}$Mpc apart (as shown in Fig.~\ref{fig:Correlation_matrix.png}). The sensitivity of our results to the inclusion of largely separated off-diagonal covariance matrix elements, demonstrate the importance of the accuracy of covariance matrix estimation.

\subsection{The Choice of Fitting Range}
\label{sec:The Choice of Fitting Range}

In order to investigate the effects of the choice of fitting range on our measured value of $\alpha$ and the significance of the detection of the BAO peak, we perform our fitting across 7 different ranges using the $\xi^{fit}$ model with and without the $A(s)$ nuisance fitting terms. We summarise the results in Table~\ref{tab:alphas_table_range}. It can be seen that the value of $\alpha$ and the magnitude of its error are largely insensitive to the choice of the fitting range for the CMASS sample. Slight variations in the value of $\alpha$ are observed as the fitting range is varied in the case of the LOWZ sample, however, these values remain consistent within the uncertainties. It can be seen that the quality of the fits produced by the $\xi^{fit}$ model without the $A(s)$ nuisance fitting terms are consistently lower than the fits produced by the complete model across various ranges as shown by the $\chi^2_{min}/dof$ values. The quantity that appears to be most sensitive to the choice of the fitting range is the significance of the detection of the BAO peak in the data. At the two extremes, the significance of the detection of the peak varies from $8.0\sigma$ to $5.3\sigma$ for the CMASS sample and from $4.3\sigma$ to $3.0\sigma$ for the LOWZ sample, depending on the choice of the fitting range. \cite{Vargas2016} have also examined the effect of the choice of fitting range on the robustness of the BAO peak measurement, reporting noisier results as the lower and upper bounds of the fitting range approach the BAO scale (i.e. $80h^{-1}$Mpc and $120h^{-1}$Mpc respectively), particularly in the former case. This level of variation highlights the importance of providing appropriate justification for the choice of fitting range in studies performing analysis of the BAO feature. 

\begin{table*}
	\centering
	\caption[Fitting the LOWZ and CMASS $\bar{\xi}(s)$ with different models over various ranges]{Results of fitting the correlation functions of the LOWZ and CMASS samples using two different models and over various fitting ranges. In performing these fits the BOSS DR12 covariance matrices were used, and as before, \lq Significance\rq\ refers to the significance of the detection of the BAO peak using the complete fitting model.}
	\label{tab:alphas_table_range}
	\begin{tabular}{c|clccc} 
		\hline
		\hline
		This Work & Range ($h^{-1}$Mpc) & Model & $\alpha$ & $\chi^2_{min}/dof$& Significance \\
		\hline
		& $28\leqslant s\leqslant180$ & $B^2\xi_m+A(s)$ &$1.0109\pm0.0121$ & 14.9/15 & $8.0\ \sigma$  \\
		&  & $B^2\xi_m$ & $1.0009\pm0.0116$ & 28.5/18 &   \\
		
		& $36\leqslant s\leqslant172$ & $B^2\xi_m+A(s)$ &$1.0134\pm0.0117$ & 11.9/13 & $6.5\ \sigma$ \\
		&  & $B^2\xi_m$ & $1.0084\pm0.0115$ & 19.0/16 &  \\
		
		& $44\leqslant s\leqslant164$ & $B^2\xi_m+A(s)$ &$1.0153\pm0.0119$ & 11.0/11 & $6.2\ \sigma$ \\
		&  & $B^2\xi_m$ & $1.0070\pm0.0117$ & 20.6/14 &   \\
		
		CMASS & $52\leqslant s\leqslant156$ & $B^2\xi_m+A(s)$ &$1.0143\pm0.0125$ & 6.6/9 & $6.5\ \sigma$ \\
		&  & $B^2\xi_m$ & $1.0070\pm0.0117$ & 23.2/12 &   \\
		
		& $60\leqslant s\leqslant148$ & $B^2\xi_m+A(s)$ &$1.0133\pm0.0123$ & 6.3/7 & $7.0\ \sigma$ \\
		&  & $B^2\xi_m$ & $1.0109\pm0.0116$ & 24.7/10 &  \\
		
		& $68\leqslant s\leqslant140$ & $B^2\xi_m+A(s)$ & $1.0148\pm0.0118$ & 6.2/5 & $7.2\ \sigma$  \\
		&  & $B^2\xi_m$ & $1.0142\pm0.0115$ & 28.3/8 &  \\
		
		& $76\leqslant s\leqslant132$ & $B^2\xi_m+A(s)$ &$1.0114\pm0.0129$ & 5.6/3 & $5.3\ \sigma$  \\
		&  & $B^2\xi_m$ & $1.0142\pm0.0109$ & 29.9/6 &  \\
		\hline
		& $28\leqslant s\leqslant180$ & $B^2\xi_m+A(s)$ &$1.0074\pm0.0266$ & 15.5/15 & $4.0\ \sigma$  \\
		&  & $B^2\xi_m$ &$0.9698\pm0.0523$ & 45.5/18 &   \\
		
		& $36\leqslant s\leqslant172$ & $B^2\xi_m+A(s)$ &$1.0121\pm0.0246$ & 13.9/13 & $4.3\ \sigma$  \\
		&  & $B^2\xi_m$ & $0.9724\pm0.0174$ & 49.0/16 &  \\
		
		& $44\leqslant s \leqslant164$ & $B^2\xi_m+A(s)$ &$1.0158\pm0.0239$ & 11.7/11 & $3.3\ \sigma$ \\
		&  & $B^2\xi_m$ & $0.9794\pm0.0174$ & 48.5/14 &  \\
		
		LOWZ & $52\leqslant s\leqslant156$ & $B^2\xi_m+A(s)$ &$1.0231\pm0.0248$ & 7.5/9 & $3.0\ \sigma$ \\
		&  & $B^2\xi_m$ & $0.9957\pm0.0187$ & 39.0/12 &  \\
		
		& $60\leqslant s\leqslant148$ & $B^2\xi_m+A(s)$ &$1.0218\pm0.0253$ & 6.9/7 & $3.2\ \sigma$ \\
		&  & $B^2\xi_m$ & $0.9949\pm0.0192$ & 43.2/10 &  \\
		
		& $68\leqslant s\leqslant140$ & $B^2\xi_m+A(s)$ &$1.0218\pm0.0250$ & 6.7/5 & $3.2\ \sigma$  \\
		&  & $B^2\xi_m$ & $0.9998\pm0.0189$ & 42.3/8 &  \\
		
		& $76\leqslant s\leqslant132$ & $B^2\xi_m+A(s)$ &$1.0303\pm0.0224$ & 5.6/3 & $3.3\ \sigma$ \\
		&  & $B^2\xi_m$ & $0.9969\pm0.0183$ & 31.3/6 &   \\
	\end{tabular}
\end{table*}

\subsection{Cosmological Distance Constraints}
\label{sec:Cosmological Distance Constraints}

Using our measured values of $\alpha$ and 5-fields $\bar{\alpha}$ presented in Table~\ref{tab:alphas_table} (for the complete $\xi^{fit}$ model), and our fiducial distances presented in Table~\ref{tab:fid_distances}, we calculate the volume-averaged distance to redshift $z$, $D_V(z)$ for the LOWZ and CMASS samples. A comparison of our results and the findings of \citet{Cuesta2016} is given in Table~\ref{tab:DV_results}. As expected given our measurements of $\alpha$, we find our results to be in agreement with those from \citet{Cuesta2016} for both samples. Furthermore, it can be seen that the magnitude of the errors are comparable between the two studies in the case of $D_V(z)$ which is based on the errors on $\alpha$ (giving a $2.6$ and $1.2$ percent distance measurement for the LOWZ and CMASS samples respectively), while the \lq5-fields $D_V(z)$\rq\ errors are larger due to the larger errors on the 5-fields $\bar{\alpha}$ values.

\begin{table*}
	\centering
	\caption[Our measurements of $D_V(z)$ in comparison with \citet{Cuesta2016}]{Distance constrains obtained from the analysis of the BAO feature in the correlation function of CMASS and LOWZ samples in this work and by \citet{Cuesta2016} (Table 11). Here $D_V(z)$ is calculated based on the value of $\alpha$ obtained from fitting to the mean correlation function of the samples, while the \lq5-fields $D_V(z)$\rq\ values are calculated based on $\bar{\alpha}$, which is obtained by taking the mean of the values of $\alpha$ attained from individually fitting to the 5 fields in the LOWZ and CMASS samples. In both cases the $\alpha$s correspond to fitting to the range $28\leqslant s \leqslant180h^{-1}$Mpc using the complete $\xi^{fit}$ model described in Eq.~\ref{eq:xi_fit}. We assume a fiducial sound horizon value of $r_{d,fid}=147.10$ Mpc. The distance constrains are quoted at the effective redshifts of $z=0.57$ and $z=0.32$ for the CMASS and LOWZ samples respectively.}
	\label{tab:DV_results}
	\begin{tabular}{l|c|c} 
		\hline
		\hline
		Study, Sample & $D_V(z)r_{d,fid}/r_d$ & 5-fields $D_V(z)r_{d,fid}/r_d$  \\
		& (Mpc) & (Mpc)	\\
		\hline
		This work, CMASS & $2031\pm24$  & $2034\pm40$  \\
		Cuesta et al. (2016), CMASS Pre-Recon & $2040\pm28$ & ----- \\
		\hline
		This work, LOWZ & $1244\pm33$  &  $1241\pm49$ \\
		Cuesta et al. (2016), LOWZ Pre-Recon & $1246\pm37$ & -----  \\
	\end{tabular}
\end{table*}

\section{Quasar BAO Analysis}
\label{sec:QSO_intro}

In this section we extend our BAO analysis to higher redshifts by
performing isotropic fitting to the combined monopole correlation
functions of four quasar samples from the 2dF QSO Redshift Survey (2QZ; \citealt{2QZ_smith2005}), SDSS Data Release 5 (SDSS DR5; \citealt{SDSS_DR5_Adelman-McCarthy2007}), 2dF-SDSS LRG and QSO survey (2SLAQ; \citealt{2SLAQ_Richards2005}) and the 2dF Quasar Dark Energy Survey pilot (2QDESp; \citealt{Chehade2016}). In total, these surveys contain $\approx80,000$ quasars in the  $0.3<z<2.2$ redshift range. As with the galaxy samples in Section~\ref{sec:Error Analysis Results}, we obtain and examine the empirical error of the combined correlation function of the QSO samples, based on the scatter in the data. 

In this work, we limit our samples to the range $0.8<z<2.2$, to allow for direct comparison and combination of our results with those from \cite{Ata2017}, who performed BAO analysis on the eBOSS survey of 147,000 quasars in this redshift range. We use the published correlation function of \cite{Ata2017} and re-fit the BAO peak for $\alpha$ using the same techniques as for our quasar sample.

\subsection{2QZ+SDSS+2SLAQ+2QDESp Datasets}
\label{sec:QSO_Datasets}
Here we provide a brief summary of the relevant properties of the quasar
samples used in our BAO analysis. A more detailed description of these
samples can be found in the referenced papers.

The 2QZ sample \citep{2QZ_cat_Croom2004} covers a total area of $721.6$
deg$^2$, containing 22,655 QSOs ($\approx31$ quasars deg$^{-2}$) up to
$z\approx 3$ in the magnitude range $18.25 <b_j< 20.85$. 

The SDSS DR5 "UNIFORM" sample was constructed by \cite{SDSS_CF_ROSS2009}
by taking a subsample of the DR5 quasar catalogue
\citep{SDSS_cat_Schneider2007}. This sample covers an area of $\approx
4000$ deg$^2$, containing 30,239 QSOs ($\approx8$ quasars deg$^{-2}$),
in the redshift range $0.3\leqslant z \leqslant 2.2$ with a magnitude
limit of $i_{SDSS}\leqslant 19.1$. 

The 2SLAQ sample \citep{2SLAQ_cat_Croom2009} covers an area of $\approx
192$ deg$^2$ containing $\approx9,000$ QSOs ($\approx47$ quasars
deg$^{-2}$) in the redshift range $z\lesssim3$ and magnitude range
$20.5<g_{SDSS}<21.85$.

The 2QDESp sample \citep{Chehade2016} covers an area of $\approx 150$
deg$^2$ in the southern sky, containing $\approx10,000$ QSOs
($\approx67$ quasars deg$^{-2}$) with magnitudes $g\leqslant22.5$. The
quasars in the sample have a mean redshift of $z=1.55$ and with $80\%$ of the objects in the sample lying in the range $0.8<z<2.5$. 

\begin{figure}
	\includegraphics[width=\columnwidth]{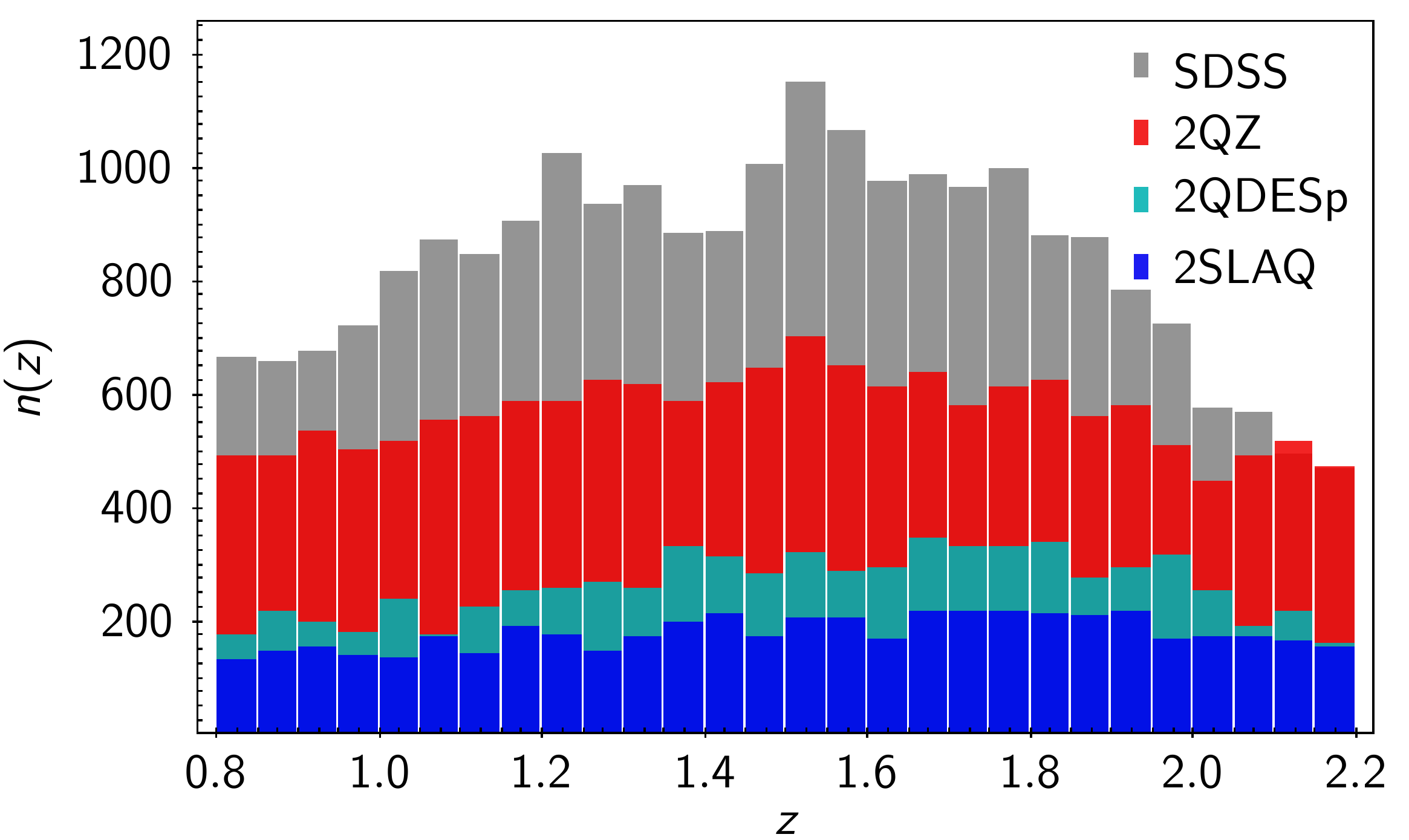}
	\caption[QSO Redshift distribution]{The redshift distribution of (from top to bottom) the SDSS DR5, 2QZ, 2QDESp and 2SLAQ QSO samples in the $0.8<z<2.2$ redshift range, analysed in this study.}
	\label{fig:QSO_nz.pdf}
\end{figure}

As mentioned above, in order to allow direct comparison and combination of our results with the measurements of \cite{Ata2017}, we restrict our analysis to objects in the redshift range $0.8<z<2.2$. This leads to a total number of quasars $N_q$, of 15,926, 23,386, 4,988 and 7,329 for the 2QZ, SDSS, 2SLAQ and 2QDESp samples respectively. Fig.~\ref{fig:QSO_nz.pdf} shows the redshift distribution $n(z)$ of the four QSO samples. The weighted mean of the correlation functions of these samples is taken to represent the correlation function of the combined quasar sample (henceforth referred to as the Combined QSO sample), containing 51,629 quasars with a mean redshift of $\bar{z}=1.5$ and an effective volume of $\approx 0.003 h^{-3}$Gpc$^3$. For comparison the QSO sample of \cite{Ata2017} covers an effective volume of $\approx 0.03 h^{-3}$Gpc$^3$, while the original SDSS LRG survey analysed by \cite{Eisenstein2005} covered an effective volume of $\approx 0.13 h^{-3}$Gpc$^3$ and the BOSS DR12 LOWZ and CMASS samples analysed by \cite{Cuesta2016} cover effective volumes of $\approx 0.67 h^{-3}$Gpc$^3$ and $\approx 1.58 h^{-3}$Gpc$^3$ respectively (in all cases we are quoting the effective volumes at $k\simeq0.15 h$Mpc$^{-1}$). 

In this section we assume the same cosmology as \cite{Ata2017} in order to facilitate direct comparison of our results, using a flat
$\Lambda$CDM cosmology with $\Omega_m=0.31$, $\Omega_bh^2=0.022$,
$h=0.676$. Although the mean redshift of the Combined QSO sample is $\bar{z}=1.5$, for simplicity and ease of comparison with \cite{Ata2017}, we quote our fiducial distance to $z=1.52$ and present our $D_V$ distance measurement at this redshift, with $D_{V,fid}(1.52)=3871.0$ Mpc and $r_{d,fid}=147.78$ Mpc. 

\subsection{eBOSS}
The eBOSS quasar survey is fully described by \cite{Ata2017}, in which BAO measurements were performed based on a sample of 147,000 quasars in the redshift 
range $0.8<z<2.2$. With an area of $\approx2044$\ deg$^2$, the quasar sky density of the sample is $\approx$72\ deg$^{-2}$. Here we simply use the correlation function from their Fig. 5 along with the QPM error bars.

\section{Measuring QSO correlation functions}
\label{sec:QSO_Methodology}      

In this section we summarise the applied methodology in our measurement of the correlation functions of the 2QDESp, 2QZ, 2SLAQ and SDSS QSO samples. We then describe our analysis of the BAO feature in our Combined QSO sample as well as the eBOSS correlation function presented by \cite{Ata2017}. 

We use the Landy-Szalay estimator (described in Section~\ref{sec:Measuring CF}), along with random catalogues generated by \cite{Chehade2016}, in order to measure the correlation functions of the four QSO samples. The random catalogues are $20\times$ larger than the data for all samples with the exception of SDSS where the random catalogue is $30\times$ larger than the data. To account for effects of photometric and spectroscopic incompleteness, \cite{Chehade2016} have applied appropriate normalisation to these randoms on a field to field basis. 

All four correlation functions are calculated using 25 $8\ h^{-1}$Mpc bins, following the same approach as our measurements of LOWZ and CMASS correlation functions in the previous sections. We found that the four individual correlation functions showed Poisson errors of varying sizes, where the Poisson error is given by $\sigma(s)=(1+\xi(s))/\sqrt{DD(s)}$. The 2QZ sample has the lowest errors, with the 2SLAQ, SDSS and 2QDESp samples having larger errors by factors of $\sim2$, $\sim1.5$ and $\sim1.5$ respectively, in our main fitting range. We therefore combined the four measured correlation functions by taking the weighted mean, given by $\hat{\xi}(s)=[\sum\xi_i(s)/\sigma_i^2(s)]/[\sum1/\sigma_i^2(s)]$, there being little difference if we combined on the basis of summing DD etc. pairs. We used the error on the weighted mean given by $\sigma_{\hat{\xi}(s)}=\sqrt{(\sum1/\sigma_i^2(s))}$ as an estimate of the error. We then fit the mean correlation function for $\alpha$ following the procedure described in Section~\ref{sec:Fitting the Correlation Function}, in the fitting range $35<s<180\ h^{-1}$Mpc. This fitting range is used when reporting our main results to match the approach in \cite{Ata2017}. However, we also fit the Combined QSO $\hat{\xi}(s)$ in the fitting range $35<s<200\ h^{-1}$Mpc in order to study any potential effects of this choice on the results, in a similar manner as in Section~\ref{sec:The Choice of Fitting Range}.

As obtaining an accurate estimation of the covariance matrix for the Combined QSO sample requires the generation of a large set of realistic mocks, a large task which lies beyond the scope of this work, when performing the fits, we simply make use of the error on the weighted mean described above. \cite{Shanks1994} have shown that the relatively low space density of quasars means that at scales up to $\sim100 h^{-1}$Mpc, the covariance between correlation function points is low. Comparing the error on the weighted mean, to the standard error on the mean (as defined in Eq.~\ref{eq:Standard_err}, providing an empirical estimate of the error), we find the two measures of the uncertainty to be close in our fitting range, with the mean ratio of empirical to Poisson error being $\sim1.2$, indicating that Poisson errors are good approximations over this range. Since Poisson only applies to  independent pair counts the expectation is that the covariances will be low. This view is partly supported by the measurements of \cite{Ata2017} in the eBOSS sample who found that the correlation between adjacent points was $\sim0.2$, with the covariance matrix being dominated by the diagonal elements. Although clearly our assumption that omission of off-diagonal terms has a negligible effect in our fits needs to be further tested. 

\section{QSO BAO Analysis: Results and Discussion}
\label{sec:QSO_Results}

\begin{figure}
	\includegraphics[width=1\linewidth]{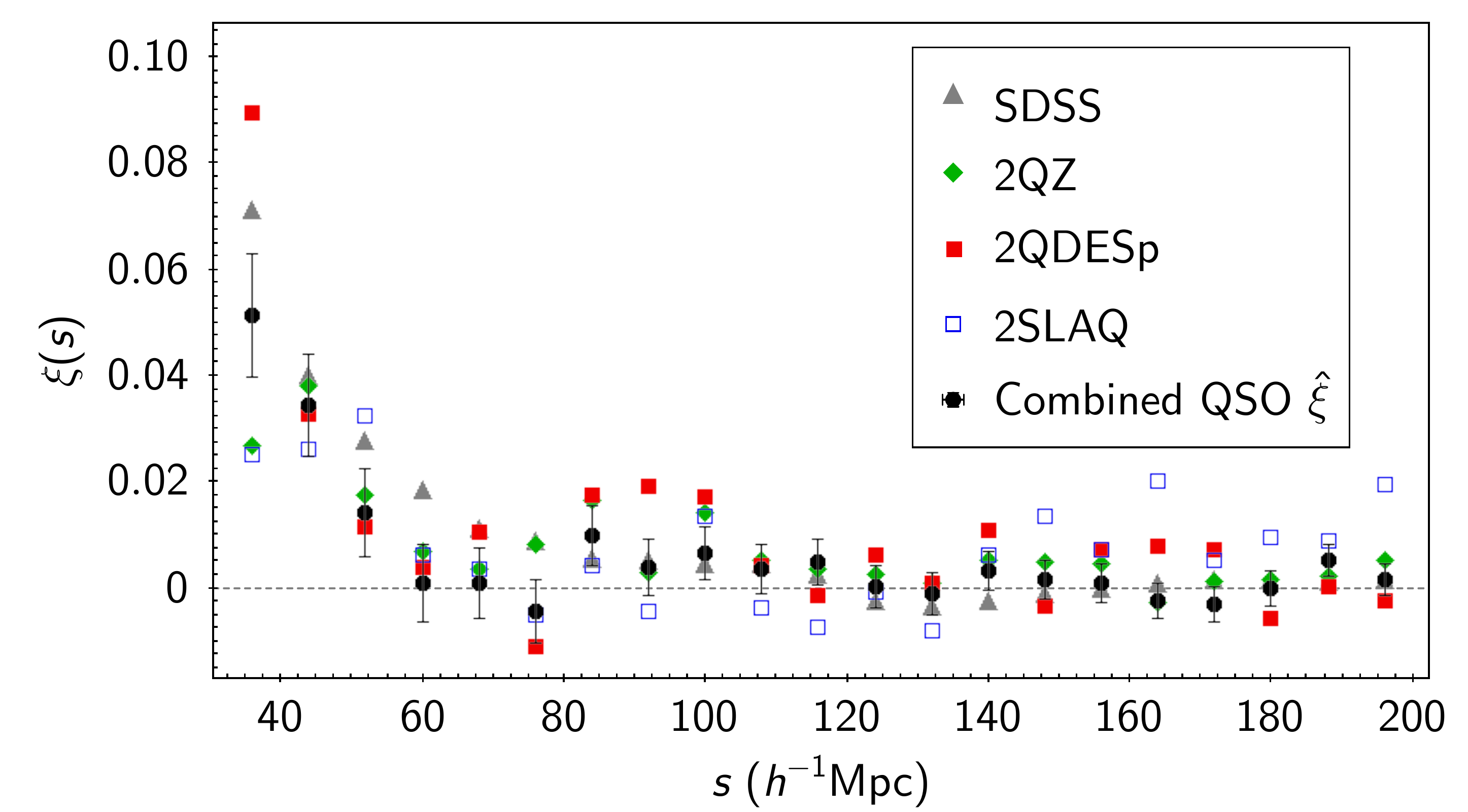}
	\caption{Correlation functions of the SDSS (grey triangles), 2QZ (green diamonds), 2QDESp (red filled squares) and 2SLAQ (blue open squares) QSO samples along with the weighted mean and the error on the weighted mean of the four samples (black). For clarity the error bars on the correlation functions of the four samples are not plotted.}
	\label{fig:QSO_CFs.pdf}
\end{figure}

In this section we present the results of our BAO analysis in the correlation
function of the Combined QSO sample as well as the eBOSS QSO correlation function of \cite{Ata2017}. The correlation functions of the SDSS, 2QZ, 2QDESp and 2SLAQ QSO samples along with the weighted mean of these correlation functions is shown in Fig.~\ref{fig:QSO_CFs.pdf}.

\subsection{Fitting the Combined QSO Sample}
\label{sec:2QZ+SDSS+2SLAQ+2QDESp}

\begin{figure}
	\begin{subfigure}{\columnwidth}
		\centering
		\caption{}
		\includegraphics[width=1\linewidth]{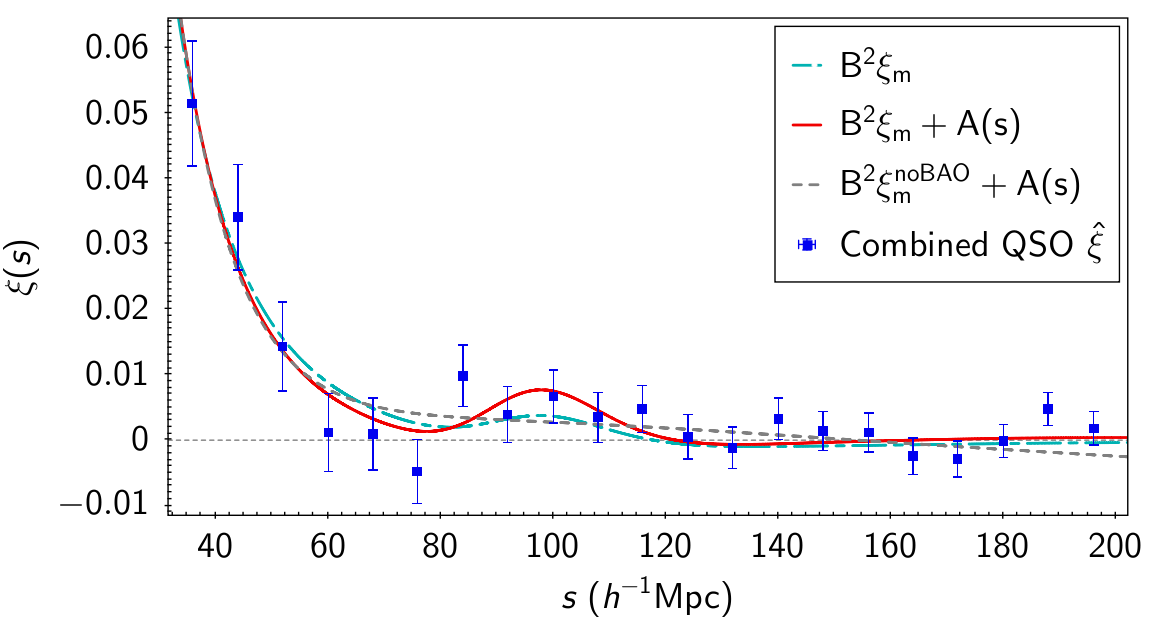}
		\label{fig:QSO_combined_fit_35-180.pdf}
	\end{subfigure}
	\begin{subfigure}{\columnwidth}
		\centering
		\caption{}
		\includegraphics[width=1\linewidth]{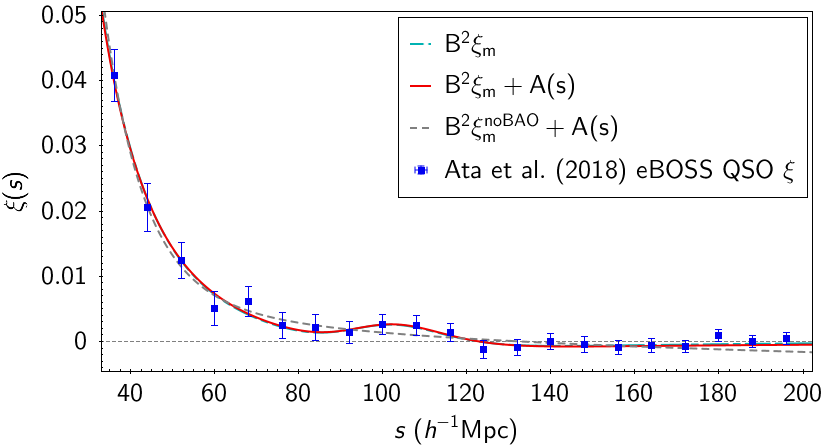}
		\label{fig:eBOSSv2_Xi_35_s_180.pdf}
	\end{subfigure}
	\begin{subfigure}{\columnwidth}
		\centering
		\caption{}
		\includegraphics[width=1\linewidth]{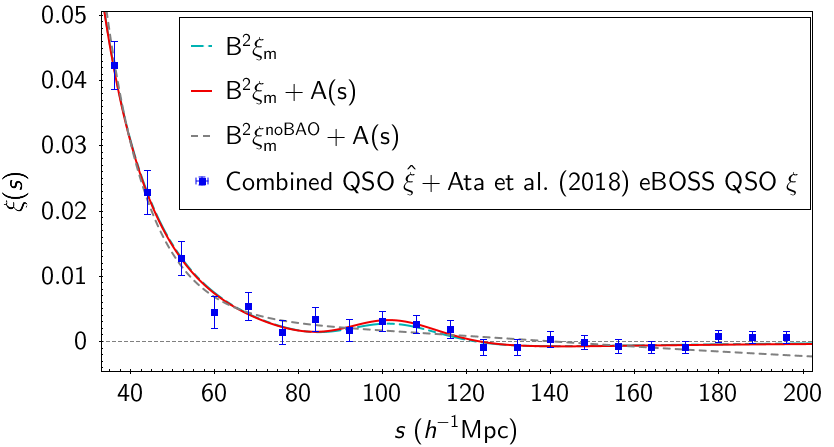}
		\label{fig:US_PE2+eBOSSv2_Xi_35_s_180.pdf}
	\end{subfigure}
	
	\caption[]{\textbf{Subplot (a):} The results of fitting the weighted mean of the 2QZ, SDSS, 2SLAQ and 2QDESp quasar samples' correlation functions. The error bars are the error on the weighted mean. The fitting is performed using various models for bins in the range $35<s<180 h^{-1}$Mpc. The dot-dashed light blue curve is the $\xi^{fit}$ model (equation \ref{eq:xi_fit}) with the $B$ fitting parameter only, while the solid red curve shows the complete $\xi^{fit}$ model. The dashed grey curve is the $\xi^{noBAO}$ model fitted with the $B$ and $A(s)$ fitting terms.
	\textbf{Subplot (b):} The eBOSS QSO correlation function taken from Fig. 5 of \cite{Ata2017} along with the QPM error bars. Also shown is our three fits to the correlation function using the same models as in subplot (a). Note that the fit given by the $B^2\xi(s)$ model (dot-dashed light blue curve) is in this case very similar to that given by our complete fitting model (solid red curve) and is therefore covered by complete model in this plot. This is also reflected in Table 5 of \cite{Ata2017}, with the fits using two models producing very similar results.
	\textbf{Subplot (c):} The weighted mean of our Combined QSO correlation function (subplot a) and the eBOSS QSO correlation function (subplot b). Fits using our three models are shown with the correlation function clearly being dominated by the eBOSS data due to its smaller error bars.}
	\label{fig:QSO_xi(s)_fits}
\end{figure}

The results of fitting to the correlation function of the combined QSO sample with the complete $\xi^{fit}$ model (equation~\ref{eq:xi_fit}), the $\xi^{fit}$ model without the $A(s)$ nuisance fitting terms, and a complete $\xi^{noBAO}$ model in the range $35<s<180 h^{-1}$Mpc, are presented in
Fig.~\ref{fig:QSO_combined_fit_35-180.pdf}. The values of $\alpha$ and $D_V(z)$
corresponding to the two variations of the $\xi^{fit}$ model are presented in Table~\ref{tab:QSO_results_table}. In contrast to fits performed in the previous chapter, upon performing an $F$-ratio test it can be seen that the complete model does not provide a significantly better fit in comparison to the simple model ($F=0.93$, $p=0.454$). For consistency with our analysis in the previous section and that of \cite{Ata2017} however, when reporting our final results, we continue to use those corresponding to the complete model. We find that fitting the correlation function in the range $35<s<200 h^{-1}$Mpc does not have a significant effect on the measurements of the BAO peak position, resulting in a $0.8\%$ shift towards larger values of $\alpha$ and a $7\%$ decrease on its uncertainty. 

\subsection{Significance of QSO BAO Peak Detection}
\label{sec:Significance of QSO BAO Peak Detection}

\begin{figure}
	\includegraphics[width=\columnwidth]{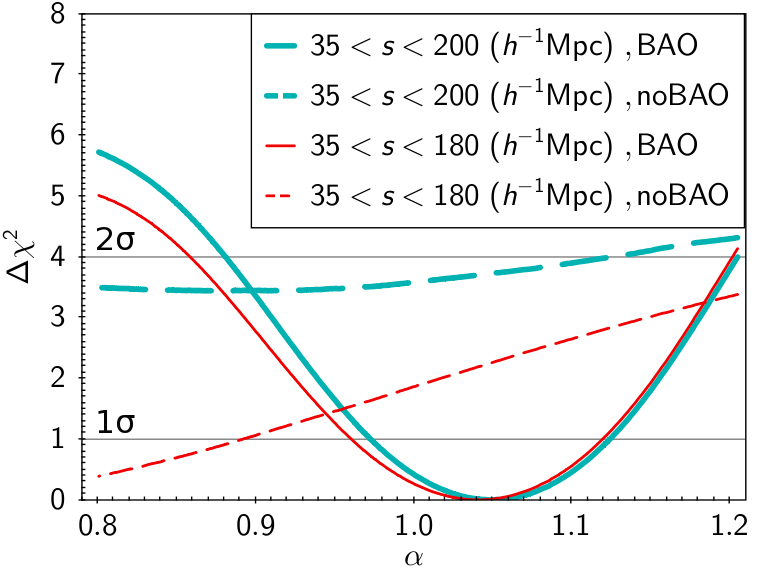}
	\caption[Significance of detection of the BAO peak in the combined QSO sample $\bar{\xi}(s)$]{The significance of the detection of the BAO peak based on fitting to the correlation function of the Combined QSO sample in the range $35<s<180h^{-1}$Mpc (red curves) and $35<s<200h^{-1}$Mpc (bold blue curves). The solid curves correspond to a fits to the data based on the $\xi^{fit}$ model which contains BAO, while the dashed curves correspond to the fits based on the $\xi^{noBAO}$ model with the BAO feature removed. In both cases the complete models including the $A(s)$ fitting terms are used. Here $\Delta\chi^2=\chi^2(\alpha)-\chi^2_{min}$, where $\chi^2_{min}$ is the minimum $\chi^2$ value using the model containing BAO. Comparing the two curves indicates that the BAO peak is detected at an $\approx1.4\sigma$ and $\approx1.9\sigma$ level in the $35<s<180h^{-1}$Mpc and $35<s<200h^{-1}$Mpc ranges respectively.}
	\label{fig:QSO_combined_delta_chi2_25-200.pdf}
\end{figure}

The $\Delta\chi^2$ curves from fitting the correlation function of the
Combined QSO sample, with the $\xi^{fit}$ and $\xi^{noBAO}$ models in our two different fitting ranges are presented in Fig.~\ref{fig:QSO_combined_delta_chi2_25-200.pdf}. A comparison of the curves shows that in the $35<s<180 h^{-1}$Mpc range, the BAO peak is detected at $\approx1.4\sigma$ in the data, while in the $35<s<200 h^{-1}$Mpc range, the peak is detected at a higher significance of $\approx1.9\sigma$. This is in line with our finding in Section~\ref{sec:The Choice of Fitting Range} where we demonstrated that the choice of the fitting range can have a large effect on the significance of detection of the BAO peak.   

\subsection{BAO Fits to eBOSS Quasar Correlation Function}
\label{sec:eBOSS}

In this section we perform a test of our BAO analysis techniques by fitting to the eBOSS QSO correlation function of \cite{Ata2017}. Fig.~\ref{fig:eBOSSv2_Xi_35_s_180.pdf} shows the eBOSS quasar DR14 correlation function of \cite{Ata2017} taken from their Fig. 5. We use the estimate with systematic weights applied (their solid line). The points are plotted in our Fig.~\ref{fig:eBOSSv2_Xi_35_s_180.pdf} as $\xi(s)$ rather than $s^2\xi(s)$ for consistency with our LRG fits. We fit these data using the same nuisance parameters as previously and, as in Section \ref{sec:2QZ+SDSS+2SLAQ+2QDESp}, we neglect the off-diagonal terms of the covariance matrix on the grounds that \cite{Ata2017} report low covariance between $\xi(s)$ points ($<0.2$). Table~\ref{tab:QSO_results_table} shows the results. We find $\alpha=1.012\pm0.051$ compared to their $\alpha=0.996\pm0.039$. Thus the estimates of $\alpha$ are similar but we report an $\approx35$\% larger error. Our $\chi^2/dof=3.0/13$ is small compared to their $\chi^2/dof=8.6/13$. Comparison against the best-fit no-BAO model (grey curve in our Fig.~\ref{fig:eBOSSv2_Xi_35_s_180.pdf}) shows only a $1.4\sigma$ detection of the BAO peak compared to $2.8\sigma$ obtained by \cite{Ata2017}. Again based on our findings in Section~\ref{sec:Significance of BAO Peak Detection}, this lower significance of detection is likely due to the fact that we are only using the error bars of \cite{Ata2017} (the square root of the diagonal elements of their covariance matrix) to perform the fitting. If so, the same behaviour as discussed in Section~\ref{sec:Significance of BAO Peak Detection}, is hinted at here. However, as the covariance matrix of \cite{Ata2017} is currently not available to us, we are unable to draw a comparison between this measurement and the peak detection significance obtained using the full matrix in a similar manner to Section~\ref{sec:Significance of BAO Peak Detection}.

With the eBOSS correlation function errors $\approx40$\% the size of those of the Combined QSO correlation function in Fig.~\ref{fig:QSO_combined_fit_35-180.pdf}, the eBOSS result is expected to  dominate the combination of these two. This is confirmed by our fits to the weighted mean of the two correlation functions shown in Fig.~\ref{fig:US_PE2+eBOSSv2_Xi_35_s_180.pdf}, and by the value of $\alpha=1.003\pm0.044$ with a significance of peak detection against the best-fit non-BAO model of $1.5\sigma$ shown in Table~\ref{tab:QSO_results_table}.

We find the errors on the correlation function of our Combined QSO sample to be $\sim2.5\times$ larger than those of eBOSS, and similarly, the error on $\alpha$ was $\sim2\times$ larger ($7.6\%$ versus $3.9\%$). This is roughly in line with the expectation as the eBOSS sample has an $\sim10\times$ larger effective volume, with errors scaling as $V_{eff}^{-1/2}$. However, we find the same $1.4\sigma$ BAO peak detection significance in our fits to the eBOSS sample and our Combined QSO sample. This is probably due to the fitted amplitude ($B^2$) for our QSO sample being unexpectedly $\sim2\times$ larger than for eBOSS, and emphasising that BAO scales can appear relatively well measured ($\sim7.6\%$) in samples where the peak is barely detectable above noise. With this caveat, if we then weigh by the respective errors on $\alpha$ as measured by us for our combined sample and by the error of \cite{Ata2017} for eBOSS, the result is $\alpha=1.005\pm0.035$. 

\subsection{BAO Distance Constraints on $D_V(z)$}
\label{sec:DV_vs_z}

\begin{table*}
	\centering
	\caption[]{(I) Results of fitting the correlation functions of the Combined QSO sample using the complete $\xi^{fit}$ model described in Eq.~\ref{eq:xi_fit} and the same model without the $A(s)$ nuisance fitting parameters, in the range $35<z<180h^{-1}$Mpc. (II) The eBOSS QSO BAO measurements presented by \cite{Ata2017}. (III) Our BAO measurements based on fitting the eBOSS QSO correlation function presented in Fig. 5 of \cite{Ata2017} along with their QPM errors. (IV) Results of our fit to the weighted mean of the combined QSO and eBOSS correlation functions. The distance constraint on $D_V$ calculated based on the measured values of $\alpha$ are included for each case. Based on our fiducial cosmology, we assume a fiducial sound horizon value of $r_{d,fid}=147.78$ Mpc. Here the value of $D_V(z)$ is quoted at $z=1.52$.}
	\label{tab:QSO_results_table}
	\begin{tabular}{l|lccc|c} 
		\hline
		\hline
		Dataset & Model & $\alpha$ & $\chi^2_{min}/dof$ & Significance & $D_V(z)r_{d,fid}/r_d$ (Mpc) \\
		\hline
		(I) Combined QSO & $B^2\xi_m+A(s)$ & $1.042\pm0.079$ & 11.6/13 & $1.4\sigma$ & $4034\pm306$\\
		& $B^2\xi_m$ & $1.033\pm0.106$ & 14.1/16& $1.8\sigma$ &  $3999\pm410$ \\	
		\hline
		(II) eBOSS QSO & $B^2\xi_m+A(s)$ & $0.996\pm0.039$ & 8.6/13 & $2.8\sigma$ & $3856\pm151$\\
		\hline		 
		(III) Our fit to eBOSS QSO & $B^2\xi_m+A(s)$ & $0.988\pm0.050$ & 3.0/13 & $1.4\sigma$ & $3825\pm194$\\
		\hline
		(IV) Combined QSO+eBOSS & $B^2\xi_m+A(s)$ & $0.997\pm0.042$ & 5.5/13 & $1.5\sigma$ & $3859\pm163$\\
	\end{tabular}	
\end{table*}

\begin{figure}
	\begin{subfigure}{\columnwidth}
		\centering
		\caption{} 
		\includegraphics[width=1\linewidth]{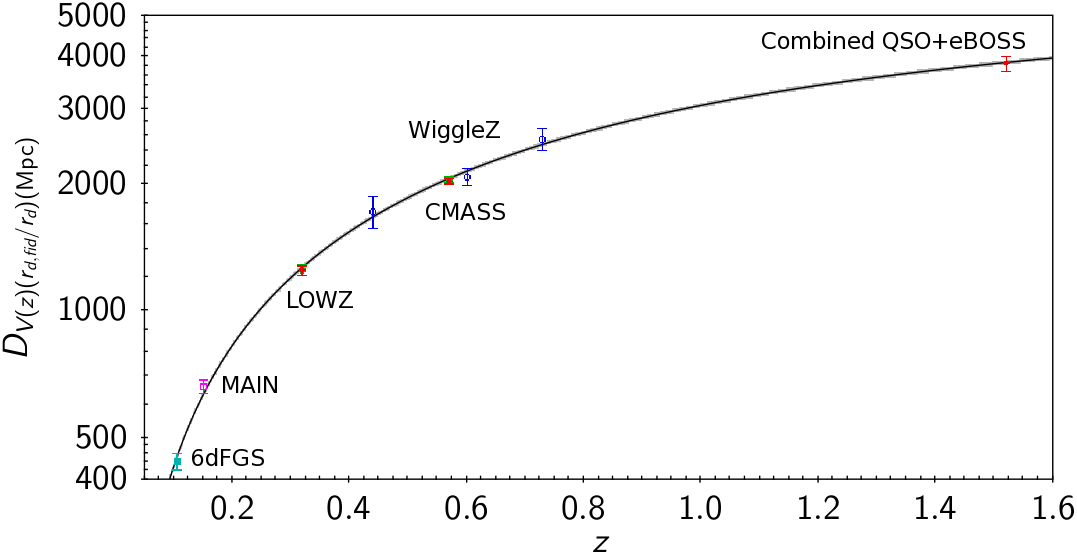}
		\label{fig:D_V_plot_log.pdf}
	\end{subfigure}	
	\begin{subfigure}{\columnwidth}
		\centering
		\caption{}
		\includegraphics[width=\columnwidth]{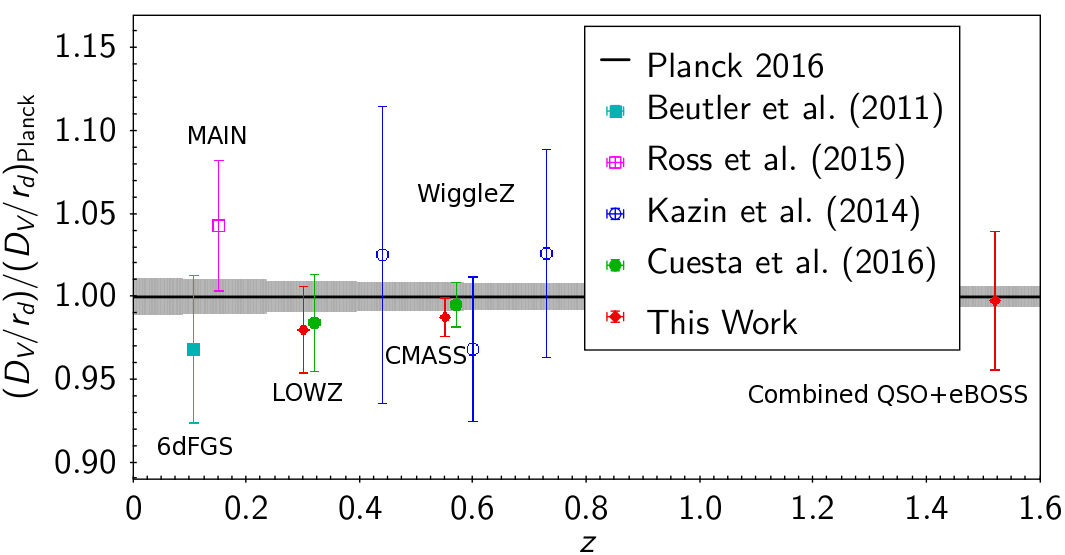}
		\label{fig:D_V_plot_ratio.pdf} 
	\end{subfigure}	
	\caption[Distance constraints on $D_V(z)$ based on various surveys in comparison to Planck 2016 predictions]{A comparison of our measured values of $D_V$ for the LOWZ, CMASS and Combined QSO+eBOSS samples (filled red diamonds) with the predictions based on a flat $\Lambda$CDM model with the Planck 2016 parameters (solid black line). The grey region represents the $1 \sigma$ variation on the Planck prediction of $D_V(z)$. The measurements of $D_V$ from \citet{Cuesta2016} for the CMASS and LOWZ samples (filled green circles), \citet{Beutler2011} for the 6dFGS sample (filled light blue square), \citet{Ross2015} for the SDSS DR7 Main sample (open pink square) and \citet{Kazin2014} for the WiggleZ galaxy sample (open dark blue circles), are also included for comparison. For easier comparison, in subplot (b) we have normalized the results to Planck 2016 and shifted our measurements for the LOWZ and CMASS samples along the x-axis.}
	\label{fig:D_V_plot.pdf}
\end{figure}

We present our measured value of $D_V$ based on our fit to the weighted mean of our Combined QSO sample and the eBOSS correlation function of \cite{Ata2017},
(Table~\ref{tab:QSO_results_table}), as well as our $D_V$ measurements for the BOSS DR12 CMASS and LOWZ samples (presented in Table~\ref{tab:DV_results} of
Section~\ref{sec:Cosmological Distance Constraints}), in Fig.~\ref{fig:D_V_plot.pdf}. The pre-reconstruction measurements of $D_V$ by \cite{Cuesta2016} based on the DR12 CMASS and LOWZ samples, as well as those from \citet{Beutler2011} for the 6dFGS sample, \citet{Ross2015} for the SDSS DR7 Main sample and \citet{Kazin2014} for the WiggleZ galaxy sample are also included for comparison. The flat $\Lambda$CDM prediction based on the Planck 2016 cosmology (TT, TE, EE+lowP+lensing+ext parameters from Table 4 of \citealt{Planck2015}) is added for comparison. The grey region represents the $1 \sigma$ variation on the Planck prediction of $D_V(z)$. As these variations are dominated by the uncertainties in $\Omega_mh^2$ (see e.g. \citealt{Andersonetal2014}), this region is determined via sampling $\Omega_mh^2$ under the assumption that it follows a Gaussian
distribution given by the Planck 2016 measurement and its $68\%$ confidence limit.

\begin{table}
	\centering
	\caption[]{Results of fitting the $\Lambda$CDM model to the $D_V(z)(r_{d,fid}/r_d)$ values plotted in Fig.~\ref{fig:D_V_plot.pdf} As there are two sets of measurements for LOWZ and CMASS, we fit to two subsets of the data  with (i) our measurements, (ii) \cite{Cuesta2016} measurements.}
	\label{tab:ED_fits_table}
	\begin{tabular}{c|cc} 
		\hline
		\hline
		&  $\Lambda$CDM+Planck (2015) & \\
		\hline
		Subset &  $\chi^2_{min}/dof$ & $p$-value \\
		\hline
		(i) &  $4.1/7$ & $0.77$  \\
		(ii) &  $2.9/7$ & $0.89$  \\
	\end{tabular}	
\end{table}

For the LOWZ, CMASS and Combined QSO+eBOSS samples, we find a good agreement between our measurement of $D_V$ and the Planck 2016 prediction. As shown by the $\chi^2_{min}/dof$ values in Table~\ref{tab:ED_fits_table}, regardless of using our measurements of $D_V$ for the LOWZ and CMASS samples or those of \cite{Cuesta2016}, overall the $\Lambda$CDM model provides a reasonably good fit to the data. Although the results appear to be over-fitted. 

\section{Conclusions}
\label{sec:Conclusions}

In this study we first obtained an independent empirical estimate of errors on the correlation functions of the BOSS DR12 LOWZ and CMASS samples. This was done by dividing each sample into subsamples, measuring the correlation functions for these fields individually and taking standard and bootstrap errors around the mean to represent the correlation function error of the entire sample. For both samples, we found general agreement between these empirical errors and those measured by \cite{Cuesta2016} from 1000 simulated DR12 QPM mocks.

Using the DR12 QPM covariance matrix of \cite{Cuesta2016}, we have obtained measurements of the position of the BAO peak based on isotropic fits, both to our mean correlation functions and to the correlation functions of 5 subsamples and taking the mean of the results. Using either method, we found our results to be in agreement with those from \cite{Cuesta2016} for both samples. Similarly, our measurement of the volume averaged distance $D_V(z)$ for both samples are in agreement with the result from \cite{Cuesta2016} and the predictions from \cite{Planck2015}.

We have demonstrated that the $A(s)$ nuisance fitting parameters play a significant role in producing a good fit, when fitting the correlation functions with a fiducial $\Lambda$CDM model. At our primary fitting range, an $F$-ratio test shows that the simple $\Lambda$CDM model without the $A(s)$ nuisance parameters is a significantly worse fit to the data compared to the full model, especially in the case of the LOWZ sample where the shape of the BAO peak appears flat to one side.  

By testing the effect of the choice of fitting range on our measurements we have further demonstrated that the measured position of the BAO peak and its uncertainty are largely insensitive to the choice of fitting range. However, the estimated significance of peak detection varies considerably depending on this choice by up to 30\% for both CMASS and LOWZ samples.

Interestingly, we observed a significant reduction in the $\chi^2_{min}/dof$ values when fitting the CMASS and LOWZ correlation functions using only the diagonal elements of the BOSS DR12 QPM covariance matrix. We mainly observed this effect in our fits where we included the nuisance parameters in the model. In these cases, the reduction in  $\chi^2_{min}/dof$ values resulted in notably lower rejections of our no-BAO model as well as a reduction in the BAO peak detection significances (from $8.0\sigma$ to $4.7\sigma$ for CMASS, and from $4.3\sigma$ to $2.8\sigma$ for LOWZ). This result shows how important the accuracy of the covariance matrix is to the determination of BAO peak significance, even at large $\approx100 h^{-1}$Mpc separations between $\xi(s)$ points.

In section~\ref{sec:QSO_intro} we extended our analysis to higher redshifts by performing fitting to the weighted mean of the correlation functions of the 2QZ, SDSS DR5, 2SLAQ and 2QDESp quasar samples. Here the BAO feature was detected at $\approx1.4\sigma$ in the data. Fitting the correlation function of our Combined QSO sample resulted in a distance constraint of $D_V(z=1.52)r_{d,fid}/r_d=4034\pm306$ Mpc (assuming $r_{d,fid}=147.78$ Mpc), a $7.6\%$ measurement to $z=1.52$. This value is in agreement with the prediction from \cite{Planck2015}, as well as the eBOSS $3.9\%$ measurement of $D_V(z=1.52)r_{d,fid}/r_d=3856\pm151$ Mpc. The main possible disagreement with the eBOSS analysis again lies in the question of the BAO peak significance since, using effectively only the diagonal elements of their covariance matrix in our fit to the eBOSS correlation function, we found a $1.4\sigma$ result (with $\chi^2_{min}/dof=3.0/13$), compared to their $2.8\sigma$ result (with $\chi^2_{min}/dof=8.6/13$), obtained using the full matrix. 

Whether we use our BAO peak results for CMASS and LOWZ or those of \cite{Cuesta2016}, there appears to be no disagreement with the standard Planck prediction for the $D_V(z)$ diagram. So once the peaks are identified, there seems little difference in the measured values of the peak positions or broadly in the errors on these positions. The main potential issue appears to be in the detection significance of the peaks which may be up to $1.7\times$ smaller than claimed in the case of CMASS LRGs and $2\times$ smaller for eBOSS quasars if only diagonal covariance matrix elements are used. Clearly our results emphasise the importance of accurate covariance matrices in correlation function analysis, even at the largest $\approx100h^{-1}$Mpc `lags' between $\xi(s)$ points. In the case of CMASS LRGs, even using our lower ($4.7\sigma$) estimate of BAO detection significance means that there is no doubt of a clear BAO detection, even before reconstruction. But for quasar samples, our lower ($1.4\sigma$) detection significance estimates mean that more data may be required to establish that the BAO peak has been unambiguously detected. It will be interesting to confirm the current quasar BAO peak detections with the full eBOSS sample and then future quasar samples from e.g. DESI.

\section*{Acknowledgements}

We would like to thank Ben Chehade, Ruari Mackenzie and Paddy Alton for their valuable advice and discussions. Additional thanks goes to Antonio J. Cuesta for kindly providing the measured correlation functions as well as the BOSS DR12 covariance matrices used in the analysis of \cite{Cuesta2016}. We would also like to thank David Alonso for providing public access to the CUTE code used in obtaining the correlation functions in this study. We would also like to thank the anonymous referee for their thorough review of this work and their helpful comments and suggestions.

Funding for SDSS-III has been provided by the Alfred P. Sloan Foundation, the Participating Institutions, the National Science Foundation, and the U.S. Department of Energy Office of Science. The SDSS-III web site is http://www.sdss3.org/.

Funding for the Sloan Digital Sky Survey IV has been provided by the Alfred P. Sloan Foundation, the U.S. Department of Energy Office of Science, and the Participating Institutions. SDSS acknowledges support and resources from the Center for High-Performance Computing at the University of Utah. The SDSS web site is www.sdss.org.

SDSS is managed by the Astrophysical Research Consortium for the Participating Institutions of the SDSS Collaboration including the Brazilian Participation Group, the Carnegie Institution for Science, Carnegie Mellon University, the Chilean Participation Group, the French Participation Group, Harvard-Smithsonian Center for Astrophysics, Instituto de Astrof\'isica de Canarias, The Johns Hopkins University, Kavli Institute for the Physics and Mathematics of the Universe (IPMU) / University of Tokyo, Lawrence Berkeley National Laboratory, Leibniz Institut f\"ur Astrophysik Potsdam (AIP), Max-Planck-Institut f\"ur Astronomie (MPIA Heidelberg), Max-Planck-Institut f\"ur Astrophysik (MPA Garching), Max-Planck-Institut f\"ur Extraterrestrische Physik (MPE), National Astronomical Observatories of China, New Mexico State University, New York University, University of Notre Dame, Observat\'orio Nacional / MCTI, The Ohio State University, Pennsylvania State University, Shanghai Astronomical Observatory, United Kingdom Participation Group, Universidad Nacional Aut\'onoma de M\'exico, University of Arizona, University of Colorado Boulder, University of Oxford, University of Portsmouth, University of Utah, University of Virginia, University of Washington, University of Wisconsin, Vanderbilt University, and Yale University.





\bibliographystyle{mnras}
\bibliography{bibliography}

\begin{thebibliography}{}
\makeatletter
\relax
\def\mn@urlcharsother{\let\do\@makeother \do\$\do\&\do\#\do\^\do\_\do\%\do\~}
\def\mn@doi{\begingroup\mn@urlcharsother \@ifnextchar [ {\mn@doi@}
  {\mn@doi@[]}}
\def\mn@doi@[#1]#2{\def\@tempa{#1}\ifx\@tempa\@empty \href
  {http://dx.doi.org/#2} {doi:#2}\else \href {http://dx.doi.org/#2} {#1}\fi
  \endgroup}
\def\mn@eprint#1#2{\mn@eprint@#1:#2::\@nil}
\def\mn@eprint@arXiv#1{\href {http://arxiv.org/abs/#1} {{\tt arXiv:#1}}}
\def\mn@eprint@dblp#1{\href {http://dblp.uni-trier.de/rec/bibtex/#1.xml}
  {dblp:#1}}
\def\mn@eprint@#1:#2:#3:#4\@nil{\def\@tempa {#1}\def\@tempb {#2}\def\@tempc
  {#3}\ifx \@tempc \@empty \let \@tempc \@tempb \let \@tempb \@tempa \fi \ifx
  \@tempb \@empty \def\@tempb {arXiv}\fi \@ifundefined
  {mn@eprint@\@tempb}{\@tempb:\@tempc}{\expandafter \expandafter \csname
  mn@eprint@\@tempb\endcsname \expandafter{\@tempc}}}

\bibitem[\protect\citeauthoryear{{Adelman-McCarthy} et~al.,}{{Adelman-McCarthy}
  et~al.}{2007}]{SDSS_DR5_Adelman-McCarthy2007}
{Adelman-McCarthy} J.~K.,  et~al., 2007, \mn@doi [ApJS] {10.1086/518864}, \href
  {http://adsabs.harvard.edu/abs/2007ApJS..172..634A} {172, 634}

\bibitem[\protect\citeauthoryear{{Alam} et~al.,}{{Alam}
  et~al.}{2015}]{Alam2015}
{Alam} S.,  et~al., 2015, \mn@doi [ApJS] {10.1088/0067-0049/219/1/12}, \href
  {http://adsabs.harvard.edu/abs/2015ApJS..219...12A} {219, 12}

\bibitem[\protect\citeauthoryear{{Alonso}}{{Alonso}}{2012}]{Alonso2012}
{Alonso} D.,  2012, preprint, \href
  {http://adsabs.harvard.edu/abs/2012arXiv1210.1833A} {} (\mn@eprint {arXiv}
  {1210.1833})

\bibitem[\protect\citeauthoryear{{Anderson} et~al.,}{{Anderson}
  et~al.}{2012}]{Andersonetal2012}
{Anderson} L.,  et~al., 2012, \mn@doi [MNRAS]
  {10.1111/j.1365-2966.2012.22066.x}, \href
  {http://adsabs.harvard.edu/abs/2012MNRAS.427.3435A} {427, 3435}

\bibitem[\protect\citeauthoryear{{Anderson} et~al.,}{{Anderson}
  et~al.}{2014}]{Andersonetal2014}
{Anderson} L.,  et~al., 2014, \mn@doi [MNRAS] {10.1093/mnras/stu523}, \href
  {http://adsabs.harvard.edu/abs/2014MNRAS.441...24A} {441, 24}

\bibitem[\protect\citeauthoryear{{Ata} et~al.,}{{Ata} et~al.}{2018}]{Ata2017}
{Ata} M.,  et~al., 2018, \mn@doi [\mnras] {10.1093/mnras/stx2630}, \href
  {http://adsabs.harvard.edu/abs/2018MNRAS.473.4773A} {473, 4773}

\bibitem[\protect\citeauthoryear{{Ballinger}, {Peacock}  \&
  {Heavens}}{{Ballinger} et~al.}{1996}]{Ballinger1996}
{Ballinger} W.~E.,  {Peacock} J.~A.,   {Heavens} A.~F.,  1996, \mn@doi [\mnras]
  {10.1093/mnras/282.3.877}, \href
  {http://adsabs.harvard.edu/abs/1996MNRAS.282..877B} {282, 877}

\bibitem[\protect\citeauthoryear{{Beutler} et~al.,}{{Beutler}
  et~al.}{2011}]{Beutler2011}
{Beutler} F.,  et~al., 2011, \mn@doi [MNRAS]
  {10.1111/j.1365-2966.2011.19250.x}, \href
  {http://adsabs.harvard.edu/abs/2011MNRAS.416.3017B} {416, 3017}

\bibitem[\protect\citeauthoryear{{Blake} \& {Glazebrook}}{{Blake} \&
  {Glazebrook}}{2003}]{Blake2003}
{Blake} C.,  {Glazebrook} K.,  2003, \mn@doi [ApJ] {10.1086/376983}, \href
  {http://adsabs.harvard.edu/abs/2003ApJ...594..665B} {594, 665}

\bibitem[\protect\citeauthoryear{{Chehade} et~al.,}{{Chehade}
  et~al.}{2016}]{Chehade2016}
{Chehade} B.,  et~al., 2016, \mn@doi [MNRAS] {10.1093/mnras/stw616}, \href
  {http://adsabs.harvard.edu/abs/2016MNRAS.459.1179C} {459, 1179}

\bibitem[\protect\citeauthoryear{{Croom}, {Smith}, {Boyle}, {Shanks}, {Miller},
  {Outram}  \& {Loaring}}{{Croom} et~al.}{2004}]{2QZ_cat_Croom2004}
{Croom} S.~M.,  {Smith} R.~J.,  {Boyle} B.~J.,  {Shanks} T.,  {Miller} L.,
  {Outram} P.~J.,   {Loaring} N.~S.,  2004, \mn@doi [MNRAS]
  {10.1111/j.1365-2966.2004.07619.x}, \href
  {http://adsabs.harvard.edu/abs/2004MNRAS.349.1397C} {349, 1397}

\bibitem[\protect\citeauthoryear{{Croom} et~al.,}{{Croom}
  et~al.}{2009}]{2SLAQ_cat_Croom2009}
{Croom} S.~M.,  et~al., 2009, \mn@doi [MNRAS]
  {10.1111/j.1365-2966.2008.14052.x}, \href
  {http://adsabs.harvard.edu/abs/2009MNRAS.392...19C} {392, 19}

\bibitem[\protect\citeauthoryear{{Cuesta} et~al.,}{{Cuesta}
  et~al.}{2016}]{Cuesta2016}
{Cuesta} A.~J.,  et~al., 2016, \mn@doi [MNRAS] {10.1093/mnras/stw066}, \href
  {http://adsabs.harvard.edu/abs/2016MNRAS.457.1770C} {457, 1770}

\bibitem[\protect\citeauthoryear{{Dawson} et~al.,}{{Dawson}
  et~al.}{2016}]{Dawson2016}
{Dawson} K.~S.,  et~al., 2016, \mn@doi [AJ] {10.3847/0004-6256/151/2/44}, \href
  {http://adsabs.harvard.edu/abs/2016AJ....151...44D} {151, 44}

\bibitem[\protect\citeauthoryear{{Delubac} et~al.,}{{Delubac}
  et~al.}{2015}]{Delubac2015}
{Delubac} T.,  et~al., 2015, \mn@doi [\aap] {10.1051/0004-6361/201423969},
  \href {http://adsabs.harvard.edu/abs/2015A%26A...574A..59D} {574, A59}

\bibitem[\protect\citeauthoryear{{Dolney}, {Jain}  \& {Takada}}{{Dolney}
  et~al.}{2006}]{Donley2006}
{Dolney} D.,  {Jain} B.,   {Takada} M.,  2006, \mn@doi [MNRAS]
  {10.1111/j.1365-2966.2005.09606.x}, \href
  {http://adsabs.harvard.edu/abs/2006MNRAS.366..884D} {366, 884}

\bibitem[\protect\citeauthoryear{{Eisenstein} \& {Hu}}{{Eisenstein} \&
  {Hu}}{1998}]{Eisenstein1998}
{Eisenstein} D.~J.,  {Hu} W.,  1998, \mn@doi [ApJ] {10.1086/305424}, \href
  {http://adsabs.harvard.edu/abs/1998ApJ...496..605E} {496, 605}

\bibitem[\protect\citeauthoryear{{Eisenstein} et~al.,}{{Eisenstein}
  et~al.}{2005}]{Eisenstein2005}
{Eisenstein} D.~J.,  et~al., 2005, \mn@doi [ApJ] {10.1086/466512}, \href
  {http://adsabs.harvard.edu/abs/2005ApJ...633..560E} {633, 560}

\bibitem[\protect\citeauthoryear{{Feldman}, {Kaiser}  \& {Peacock}}{{Feldman}
  et~al.}{1994}]{FKP}
{Feldman} H.~A.,  {Kaiser} N.,   {Peacock} J.~A.,  1994, \mn@doi [ApJ]
  {10.1086/174036}, \href {http://cdsads.u-strasbg.fr/abs/1994ApJ...426...23F}
  {426, 23}

\bibitem[\protect\citeauthoryear{{Glazebrook} \& {Blake}}{{Glazebrook} \&
  {Blake}}{2005}]{Glazebrook2005}
{Glazebrook} K.,  {Blake} C.,  2005, \mn@doi [ApJ] {10.1086/432497}, \href
  {http://adsabs.harvard.edu/abs/2005ApJ...631....1G} {631, 1}

\bibitem[\protect\citeauthoryear{Gregory}{Gregory}{2005}]{Gregory2005}
Gregory P.,  2005, Bayesian Logical Data Analysis for the Physical Sciences.
Cambridge University Press, New York, NY, USA

\bibitem[\protect\citeauthoryear{{Kazin} et~al.,}{{Kazin}
  et~al.}{2014}]{Kazin2014}
{Kazin} E.~A.,  et~al., 2014, \mn@doi [MNRAS] {10.1093/mnras/stu778}, \href
  {http://adsabs.harvard.edu/abs/2014MNRAS.441.3524K} {441, 3524}

\bibitem[\protect\citeauthoryear{{Landy} \& {Szalay}}{{Landy} \&
  {Szalay}}{1993}]{LS1993}
{Landy} S.~D.,  {Szalay} A.~S.,  1993, \mn@doi [ApJ] {10.1086/172900}, \href
  {http://adsabs.harvard.edu/abs/1993ApJ...412...64L} {412, 64}

\bibitem[\protect\citeauthoryear{{Lewis}, {Challinor}  \& {Lasenby}}{{Lewis}
  et~al.}{2000}]{Lewis2000}
{Lewis} A.,  {Challinor} A.,   {Lasenby} A.,  2000, \mn@doi [ApJ]
  {10.1086/309179}, \href {http://adsabs.harvard.edu/abs/2000ApJ...538..473L}
  {538, 473}

\bibitem[\protect\citeauthoryear{{Linder}}{{Linder}}{2003}]{Linder2003}
{Linder} E.~V.,  2003, \mn@doi [\prd] {10.1103/PhysRevD.68.083504}, \href
  {http://adsabs.harvard.edu/abs/2003PhRvD..68h3504L} {68, 083504}

\bibitem[\protect\citeauthoryear{Matsubara}{Matsubara}{2004}]{Matsubara2004}
Matsubara T.,  2004, The Astrophysical Journal, 615, 573

\bibitem[\protect\citeauthoryear{{Norberg}, {Baugh}, {Gazta{\~n}aga}  \&
  {Croton}}{{Norberg} et~al.}{2009}]{Norberg2009}
{Norberg} P.,  {Baugh} C.~M.,  {Gazta{\~n}aga} E.,   {Croton} D.~J.,  2009,
  \mn@doi [MNRAS] {10.1111/j.1365-2966.2009.14389.x}, \href
  {http://adsabs.harvard.edu/abs/2009MNRAS.396...19N} {396, 19}

\bibitem[\protect\citeauthoryear{{Peebles}}{{Peebles}}{1980}]{Peebles1980}
{Peebles} P.~J.~E.,  1980, {The large-scale structure of the universe}

\bibitem[\protect\citeauthoryear{{Peebles} \& {Ratra}}{{Peebles} \&
  {Ratra}}{2003}]{Peebles2003}
{Peebles} P.~J.,  {Ratra} B.,  2003, \mn@doi [Reviews of Modern Physics]
  {10.1103/RevModPhys.75.559}, \href
  {http://adsabs.harvard.edu/abs/2003RvMP...75..559P} {75, 559}

\bibitem[\protect\citeauthoryear{{Perlmutter} et~al.,}{{Perlmutter}
  et~al.}{1999}]{Perlmutter1999}
{Perlmutter} S.,  et~al., 1999, \mn@doi [ApJ] {10.1086/307221}, \href
  {http://adsabs.harvard.edu/abs/1999ApJ...517..565P} {517, 565}

\bibitem[\protect\citeauthoryear{{Planck Collaboration} et~al.,}{{Planck
  Collaboration} et~al.}{2016}]{Planck2015}
{Planck Collaboration} et~al., 2016, \mn@doi [\aap]
  {10.1051/0004-6361/201525830}, \href
  {http://adsabs.harvard.edu/abs/2016A%26A...594A..13P} {594, A13}

\bibitem[\protect\citeauthoryear{{Reid} et~al.,}{{Reid}
  et~al.}{2016}]{Reid2016}
{Reid} B.,  et~al., 2016, \mn@doi [MNRAS] {10.1093/mnras/stv2382}, \href
  {http://adsabs.harvard.edu/abs/2016MNRAS.455.1553R} {455, 1553}

\bibitem[\protect\citeauthoryear{{Richards} et~al.,}{{Richards}
  et~al.}{2005}]{2SLAQ_Richards2005}
{Richards} G.~T.,  et~al., 2005, \mn@doi [MNRAS]
  {10.1111/j.1365-2966.2005.09096.x}, \href
  {http://adsabs.harvard.edu/abs/2005MNRAS.360..839R} {360, 839}

\bibitem[\protect\citeauthoryear{{Riess} et~al.,}{{Riess}
  et~al.}{1998}]{Riess1998}
{Riess} A.~G.,  et~al., 1998, \mn@doi [AJ] {10.1086/300499}, \href
  {http://adsabs.harvard.edu/abs/1998AJ....116.1009R} {116, 1009}

\bibitem[\protect\citeauthoryear{{Ross} et~al.,}{{Ross}
  et~al.}{2009}]{SDSS_CF_ROSS2009}
{Ross} N.~P.,  et~al., 2009, \mn@doi [ApJ] {10.1088/0004-637X/697/2/1634},
  \href {http://adsabs.harvard.edu/abs/2009ApJ...697.1634R} {697, 1634}

\bibitem[\protect\citeauthoryear{{Ross}, {Samushia}, {Howlett}, {Percival},
  {Burden}  \& {Manera}}{{Ross} et~al.}{2015}]{Ross2015}
{Ross} A.~J.,  {Samushia} L.,  {Howlett} C.,  {Percival} W.~J.,  {Burden} A.,
  {Manera} M.,  2015, \mn@doi [MNRAS] {10.1093/mnras/stv154}, \href
  {http://adsabs.harvard.edu/abs/2015MNRAS.449..835R} {449, 835}

\bibitem[\protect\citeauthoryear{{Ross} et~al.,}{{Ross}
  et~al.}{2017}]{Ross2016}
{Ross} A.~J.,  et~al., 2017, \mn@doi [\mnras] {10.1093/mnras/stw2372}, \href
  {http://adsabs.harvard.edu/abs/2017MNRAS.464.1168R} {464, 1168}

\bibitem[\protect\citeauthoryear{{S{\'a}nchez}, {Baugh}  \&
  {Angulo}}{{S{\'a}nchez} et~al.}{2008}]{Sanchez2008}
{S{\'a}nchez} A.~G.,  {Baugh} C.~M.,   {Angulo} R.~E.,  2008, \mn@doi [MNRAS]
  {10.1111/j.1365-2966.2008.13769.x}, \href
  {http://adsabs.harvard.edu/abs/2008MNRAS.390.1470S} {390, 1470}

\bibitem[\protect\citeauthoryear{{Sawangwit}, {Shanks}, {Croom}, {Drinkwater},
  {Fine}, {Parkinson}  \& {Ross}}{{Sawangwit} et~al.}{2012}]{Sawangwit2012}
{Sawangwit} U.,  {Shanks} T.,  {Croom} S.~M.,  {Drinkwater} M.~J.,  {Fine} S.,
  {Parkinson} D.,   {Ross} N.~P.,  2012, \mn@doi [\mnras]
  {10.1111/j.1365-2966.2011.19848.x}, \href
  {http://adsabs.harvard.edu/abs/2012MNRAS.420.1916S} {420, 1916}

\bibitem[\protect\citeauthoryear{{Schneider} et~al.,}{{Schneider}
  et~al.}{2007}]{SDSS_cat_Schneider2007}
{Schneider} D.~P.,  et~al., 2007, \mn@doi [AJ] {10.1086/518474}, \href
  {http://adsabs.harvard.edu/abs/2007AJ....134..102S} {134, 102}

\bibitem[\protect\citeauthoryear{{Seo} \& {Eisenstein}}{{Seo} \&
  {Eisenstein}}{2003}]{Seo2003}
{Seo} H.-J.,  {Eisenstein} D.~J.,  2003, \mn@doi [ApJ] {10.1086/379122}, \href
  {http://adsabs.harvard.edu/abs/2003ApJ...598..720S} {598, 720}

\bibitem[\protect\citeauthoryear{{Shanks}}{{Shanks}}{1985}]{Shanks1985}
{Shanks} T.,  1985, \mn@doi [Vistas in Astronomy]
  {10.1016/0083-6656(85)90062-5}, \href
  {http://adsabs.harvard.edu/abs/1985VA.....28..595S} {28, 595}

\bibitem[\protect\citeauthoryear{{Shanks} \& {Boyle}}{{Shanks} \&
  {Boyle}}{1994}]{Shanks1994}
{Shanks} T.,  {Boyle} B.~J.,  1994, \mn@doi [\mnras] {10.1093/mnras/271.4.753},
  \href {http://adsabs.harvard.edu/abs/1994MNRAS.271..753S} {271, 753}

\bibitem[\protect\citeauthoryear{{Slosar} et~al.,}{{Slosar}
  et~al.}{2013}]{Slosar2013}
{Slosar} A.,  et~al., 2013, \mn@doi [\jcap] {10.1088/1475-7516/2013/04/026},
  \href {http://adsabs.harvard.edu/abs/2013JCAP...04..026S} {4, 026}

\bibitem[\protect\citeauthoryear{{Smith}, {Croom}, {Boyle}, {Shanks}, {Miller}
  \& {Loaring}}{{Smith} et~al.}{2005}]{2QZ_smith2005}
{Smith} R.~J.,  {Croom} S.~M.,  {Boyle} B.~J.,  {Shanks} T.,  {Miller} L.,
  {Loaring} N.~S.,  2005, \mn@doi [MNRAS] {10.1111/j.1365-2966.2005.08870.x},
  \href {http://adsabs.harvard.edu/abs/2005MNRAS.359...57S} {359, 57}

\bibitem[\protect\citeauthoryear{{Squires}}{{Squires}}{2001}]{Squires}
{Squires} G.~L.,  2001, Practical Physics, 4th edn.
Cambridge University Press, Cambridge

\bibitem[\protect\citeauthoryear{{Vargas-Maga{\~n}a}
  et~al.,}{{Vargas-Maga{\~n}a} et~al.}{2016}]{Vargas2016}
{Vargas-Maga{\~n}a} M.,  et~al., 2016, preprint, \href
  {http://adsabs.harvard.edu/abs/2016arXiv161003506V} {} (\mn@eprint {arXiv}
  {1610.03506})

\bibitem[\protect\citeauthoryear{{White}, {Tinker}  \& {McBride}}{{White}
  et~al.}{2014}]{WTM2014}
{White} M.,  {Tinker} J.~L.,   {McBride} C.~K.,  2014, \mn@doi [MNRAS]
  {10.1093/mnras/stt2071}, \href
  {http://adsabs.harvard.edu/abs/2014MNRAS.437.2594W} {437, 2594}

\bibitem[\protect\citeauthoryear{{Xu}, {Padmanabhan}, {Eisenstein}, {Mehta}  \&
  {Cuesta}}{{Xu} et~al.}{2012}]{Xu2012}
{Xu} X.,  {Padmanabhan} N.,  {Eisenstein} D.~J.,  {Mehta} K.~T.,   {Cuesta}
  A.~J.,  2012, \mn@doi [MNRAS] {10.1111/j.1365-2966.2012.21573.x}, \href
  {http://adsabs.harvard.edu/abs/2012MNRAS.427.2146X} {427, 2146}

\makeatother
\end{thebibliography}




\appendix
\label{sec:Appendix}
\section{CMASS 30 Fields}
\label{sec:CMASS 30 Fields}

\begin{figure}
	\includegraphics[width=1\columnwidth]{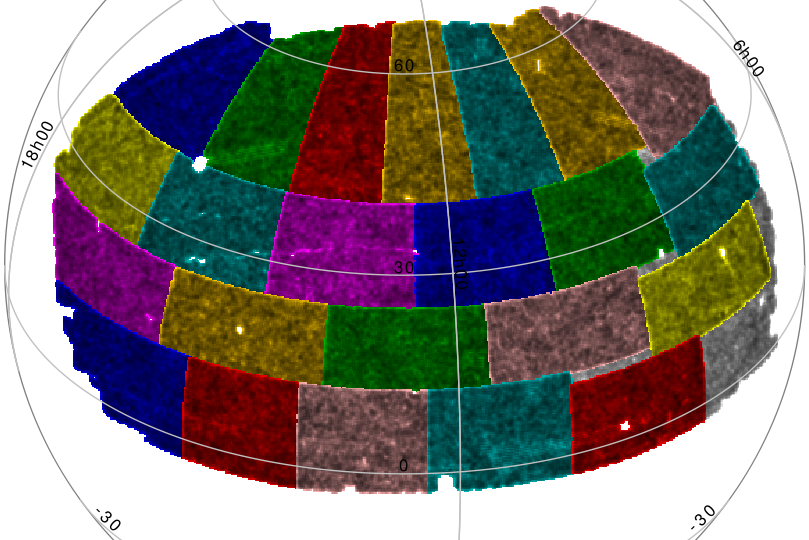}
	\label{fig:30_NGC.pdf}
	\includegraphics[width=1\columnwidth]{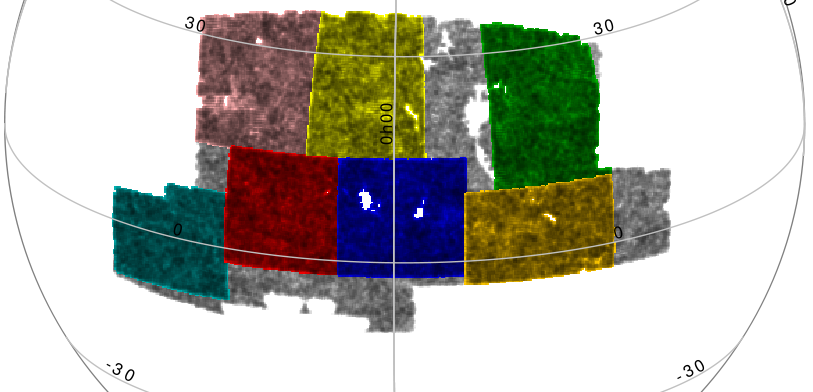}
	\label{fig:30_SGC.pdf}
	\captionof{figure}{The coverage of the chosen 30 fields in the Northern (top) and Southern (bottom) Galactic caps of the CMASS sample. The selected fields are highlighted by various colours while the unselected areas are shown in grey. These subsamples are used in order to obtain a a more accurate bootstrap estimation of the correlation function errors from the data.}
	\label{fig:CMASS_30_fields}
\end{figure}

\begin{figure}
	\includegraphics[width=1\linewidth]{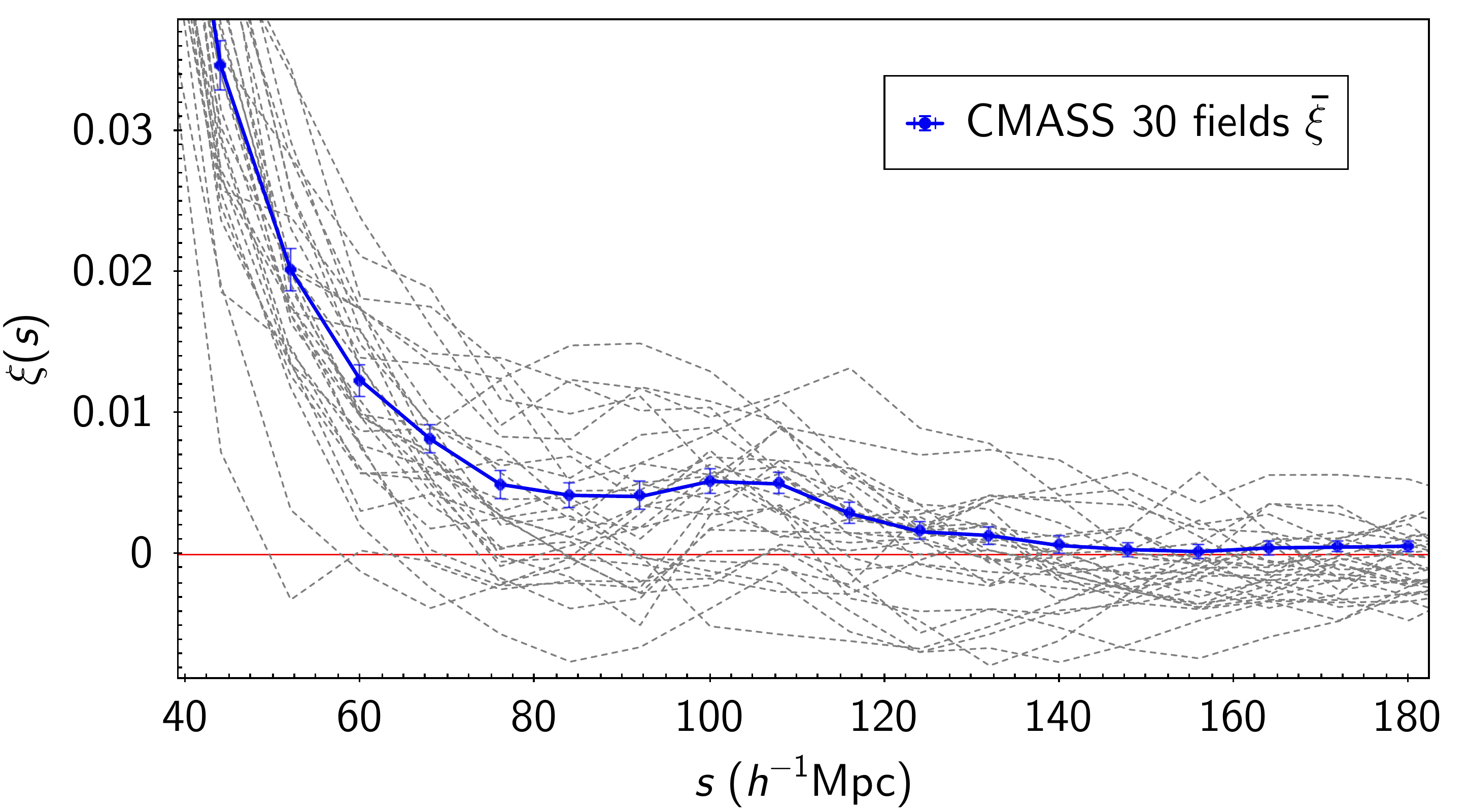}
	\captionof{figure}{Correlation functions of the 30 fields in the CMASS sample (grey dashed lines) and the mean correlation function (solid blue line). The error bars on the mean correlation function are the standard error on the mean.}
	\label{fig:CMASS_30_fields_cfs}
\end{figure}

Here we present a brief comparison of the errors achieved using the original 5 subsamples and an increased number of subsamples (30), to test the robustness of our estimated errors to subsample size. The position of the 30 selected fields are shown in Fig.~\ref{fig:CMASS_30_fields} and the corresponding correlation functions in Fig.~\ref{fig:CMASS_30_fields_cfs}. Each field contains about 23,500 galaxies and has an area of $\simeq275 \deg^2$, with the selected fields covering $88\%$ of the total sample area. We find the mean correlation function to be in a good agreement with the mean correlation function from our 5 fields as well as the CMASS correlation function from \cite{Cuesta2016}, once integral constraint (as discussed in \citealt{Peebles1980}) is accounted for. We estimate the bootstrap error on the CMASS correlation function based on these 30 subsamples and compare the results with our errors based on the original 5 subsamples in Fig.~\ref{fig:CMASS_errors.pdf}

\section{CMASS correlation matrix}
\label{sec:CMASS correlation matrix}

\begin{figure*}
	\includegraphics[width=2\columnwidth]{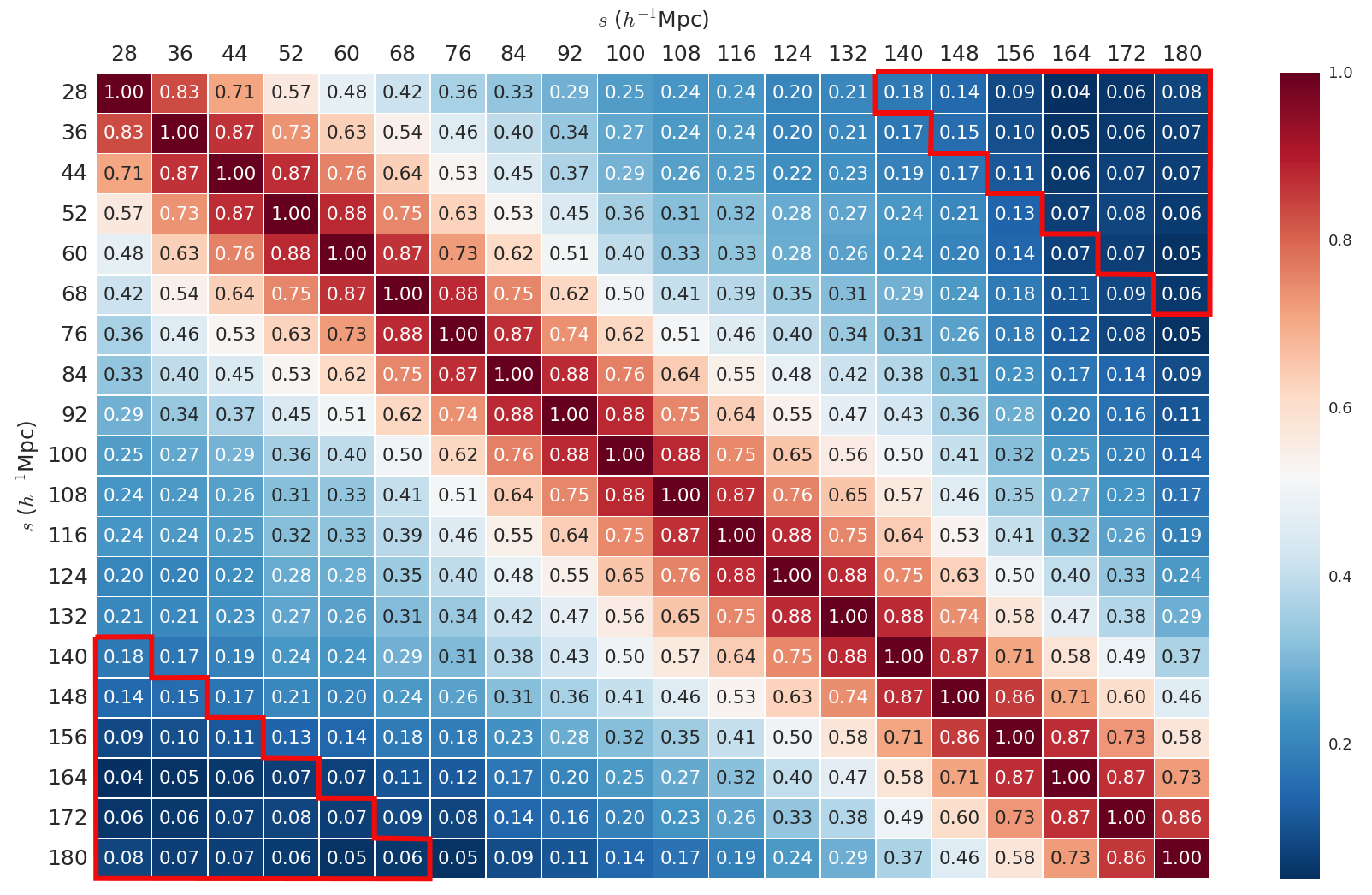}
	\captionof{figure}{CMASS DR12 Correlation matrix based on the covariance matrix used in the analysis of \citet{Cuesta2016}, as well as in our fits in this work. The red outlines indicates the 15-20 off-diagonal elements corresponding to covariance matrix terms which appear to be essential in obtaining a reasonable $\chi^2_{min}\approx15$, as shown in our test in Section~\ref{sec:Significance of BAO Peak Detection}.}
	\label{fig:Correlation_matrix.png}
\end{figure*}


\bsp	
\label{lastpage}
\end{document}